\documentclass{emulateapj}
\usepackage{epsfig,natbib,amsmath}

\newcommand{\mic}{\,$\mu$m }
\newcommand{\micpa}{\,$\mu$m}          
\newcommand{\muJy}{\,$\mu$Jy }
\newcommand{\muJypa}{\,$\mu$Jy}                              
\newcommand{\spi}{{\it Spitzer}}
\newcommand{\Lsol}{L$_\odot$}
\newcommand{\Msol}{M$_\odot$}
\newcommand{\Lir}{L$_{\rm IR}$}

\bibpunct[]{(}{)}{;}{a}{}{,}
 
\shorttitle{{\it Spitzer} observations of GRB host galaxies}
\shortauthors{E.\,Le Floc'h et al.}

\begin{document}
\def\gtapp
{\mathrel{\hbox{\raise0.3ex\hbox{$>$}\kern-0.8em\lower0.8ex\hbox{$\sim$}}}}
\def\ltapp
{\mathrel{\hbox{\raise0.3ex\hbox{$<$}\kern-0.75em\lower0.8ex\hbox{$\sim$}}}}
\def\ts{\thinspace}

\title{Probing the cosmic star formation using long Gamma-Ray
Bursts: New constraints from the Spitzer Space Telescope\footnote{Based
on observations made with {\it Spitzer}, operated by the Jet
Propulsion Laboratory under NASA contract 1407.}  }

\slugcomment{Accepted for publication in  The Astrophysical Journal}

\author{Emeric~Le~Floc'h\altaffilmark{1,2},
Vassilis~Charmandaris\altaffilmark{3,4,2},
William~J.~Forrest\altaffilmark{5},
F\'elix~Mirabel\altaffilmark{6},
Lee~Armus\altaffilmark{7}, and Daniel~Devost\altaffilmark{4}}

\altaffiltext{1}{Steward Observatory, University of Arizona, 933  North Cherry Avenue, Tucson, AZ 85721, USA}
\altaffiltext{2}{Chercheur Associ\'e, Observatoire de Paris, F-75014,  Paris, France}
\altaffiltext{3}{Department of Physics, University of Crete,  GR-71003, Heraklion, Greece}
\altaffiltext{4}{Center for Radiophysics and Space Research, Cornell University,
Space Sciences Building, Ithaca, NY 14853-6801, USA}
\altaffiltext{5}{Dept. of Physics \& Astronomy, University of Rochester,  Rochester, NY 14627, USA}
\altaffiltext{6}{European Southern Observatory, Alonso de C\'ordova  3107, Vitacura, Casilla 19001, Santiago 19, Chile}
\altaffiltext{7}{Spitzer Science Center, California Institute
of Technology, 220-6, Pasadena, CA 91125, USA}

\email{elefloch@mips.as.arizona.edu,vassilis@physics.uoc.gr,\\
forrest@pas.rochester.edu,fmirabel@eso.org,lee@ipac.caltech.edu,\\
devost@astro.cornell.edu}

\begin{abstract} 
  We report on IRAC-4.5\micpa, IRAC-8.0\mic and MIPS-24\mic deep
  observations of 16 Gamma-Ray Burst (GRBs) host galaxies performed
  with the \spi \, {\it Space Telescope,} and we investigate in the
  thermal infrared the presence of evolved stellar populations and
  dust-enshrouded star-forming activity associated with these objects.
  Our sample is derived from GRBs that were identified with sub-arcsec
  localization between 1997 and 2001, and only a very small fraction
  ($\sim$\,20\%) of the targeted sources is detected down to $f_{4.5
    \mu m}$\,$\sim$\,3.5\muJy and $f_{24 \mu
    m}$\,$\sim$\,85\muJypa~(3$\sigma$). This likely argues against a
  population dominated by massive and strongly-starbursting (i.e.,
  SFR\,$\gtapp$\,100\,\Msol\,yr$^{-1}$) galaxies as it has been
  recently suggested from submillimeter/radio and optical studies of
  similarly-selected GRB hosts.  Furthermore we find evidence that
  some GRBs do not occur in the most infrared-luminous regions --
  hence the most actively star-forming environments -- of their host
  galaxies.  Should the GRB hosts be representative of all
  star-forming galaxies at high redshift, models of infrared galaxy
  evolution indicate that $\gtapp$\,50\% of GRB hosts should have
  $f_{24 \mu m}$\,$\gtapp$\,100\muJypa. Unless the identification of
  GRBs prior to 2001 was prone to strong selection effects biasing our
  sample against dusty galaxies, we infer in this context that the
  GRBs identified with the current techniques can not be directly used
  as unbiased probes of the global and integrated star formation
  history of the Universe.

\end{abstract}

\keywords{ 
  galaxies: high-redshift ---  
  infrared: galaxies ---
  cosmology: observations --- 
  galaxies: individual (GRB970828, GRB980425, GRB980613, GRB980703, GRB981226, GRB990705)
}

\section{Introduction}

It is now widely believed that the so-called ``long'' Gamma-Ray Bursts
(i.e., GRBs with duration $\gtapp$\,2s and soft spectra, as opposed to
the short and hard bursts, e.g., \citealt{Kouveliotou93}) are
intimately connected to the collapse and the cataclismic destruction
of some short-lived and very massive stars.  Evidence in this regard
include (i) the signature of Type Ic supernova and hypernova in the
optical transient emission of GRB counterparts
\citep[e.g.,][]{Galama98,Stanek03,Hjorth03, Malesani04}; (ii) heavy
elements from metal-enriched media typical of supernova remnants
observed in the spectrum of several X-ray afterglows
\citep{Piro00,Reeves02}; (iii) the unquestionable starbursting nature
of their host galaxies
\citep[e.g.,][]{Bloom98a,Fruchter99a,Sokolov01,Chary02,LeFloch03,Christensen04};
and (iv) the location of GRBs relative to the center of their hosts,
which appears to be consistent with a population of progenitors
residing in galaxy disks \citep{Bloom02a}. Hence it has often been
proposed that long GRBs could be used as powerful tracers of the
global star-forming activity in the early Universe
\citep[e.g.,][]{Wijers98,Mirabel00,Blain00}. As illustrated by
the recent identification of a burst at $z$\,$=$\,6.29
\citep[e.g.,][]{Kawai05}, GRBs are indeed
detectable up to very high redshifts \citep{Lamb00}. They are also
very little affected by dust extinction, which is known to be
particularly significant in distant starburst galaxies
\citep[e.g.,][]{Blain99,Franceschini01,Chary01}.

However, this picture relies on the strong assumption that the
production of GRBs in a given starburst region scales only with the
rate of stars formed in this environment, with no dependency on other
physical parameters that may vary from one galaxy to another or that
may evolve throughout the lifetime of the Universe.  This assumption
is rather major, as the occurrence of a GRB could strongly depend on
the properties of the gas from which its progenitor originates (e.g.,
metallicity: \citealt{MacFadyen99,Ramirez_Ruiz02b,Heger03,Hirschi05}).  
It could also vary according to the fraction of those progenitors 
involved in binary systems \citep{Izzard04,Podsiadlowski04,Mirabel04b}, and it 
finally relies
on a non-evolution of the Initial Mass Function with redshift.
Testing this ``one-to-one'' connection between GRBs and star formation
is therefore particularly crucial to guarantee an accurate
understanding of the use of GRBs as quantitative tracers of galaxy
evolution.

One possible approach to investigate this relation is to compare the
properties of the hosts of Gamma-Ray Bursts with respect to the
galaxies responsible for the bulk of the star-forming activity in the
Universe as a function of redshift.  It is now well established that
a significant fraction
 of the present-day stellar mass budget was formed during brief
and violent infrared-luminous (L$_{\rm IR}$\,=\,L$_{\rm 8-1000\mu
  m}$\,$\gtapp$\,10$^{11}$\,\Lsol) episodes of star formation within
massive ($\mathcal{M}$\,$\gtapp$\,5\,$\times$\,10$^{10}$\,M$_\odot$)
galaxies at 0.5\,$\ltapp$\,$z$\,$\ltapp$\,3
\citep[e.g.,][]{Flores99,Blain02,Elbaz02,Dickinson03,Lagache03,Franceschini03,LeFloch05,Hammer05}.
Investigating the extinction-corrected star formation rate and the
mass of GRB hosts should therefore provide tight constraints on the
relevance of GRBs for probing the star formation history of
the Universe.

Previous observational studies on GRB hosts though have led to conflicting
views about their nature.  Based on their properties at optical
and near-infrared wavelengths it has been argued that GRB hosts are
mostly blue, sub-luminous and low-mass galaxies with young stellar
populations, characterized by a modest activity of star formation and
potential selection effects due to low metallicity
\citep[e.g.,][]{Sokolov01,LeFloch03,Fynbo03,Courty04,Prochaska04,Christensen04}. On
the other hand it has also been claimed that the morphology of these
objects and their average radio/submillimeter properties rather
indicate massive and actively star-forming galaxies
\citep{Conselice05,Berger03}.  In order to bring tighter
constraints on the nature of those sources, we undertook a survey
of GRB host galaxies with the \spi \, {\it Space Telescope.} In this
paper we present mid-infrared (mid-IR) images of 16 objects at
4.5\micpa, 8\mic and 24\micpa, which allows us to constrain the
presence of evolved stellar populations and dust-enshrouded
star-forming activity in these sources. In Sect.\,2 we describe the
\spi \, data used in this study. Sect.\,3 outlines general results
derived from these mid-IR observations, while the GRB host properties
are detailed on a galaxy-by-galaxy basis in Sect.\,4.  Our findings
are discussed in Sect.\,5, and they are finally summarized in
Sect.\,6.  Throughout this work, we assume a $\Lambda$CDM cosmology
with H$_0$\,=\,70~km~s$^{-1}$\,Mpc$^{-1}$, $\Omega_m$\,=\,0.3 and
$\Omega_{\lambda}\,=\,0.7$ \citep{Spergel03}.

\section{Observations}

\subsection{The data}

With the exception of the GRB\,970228 which was detected in a
sub-luminous blue dwarf galaxy at $z$\,=\,0.695 \citep{Bloom01b}, our
sample of GRB host galaxies was built by considering every GRB
identified before July 10$^{\rm th}$, 1999 and localized on the sky
with a sub-arscsec accuracy thanks to the detection of an afterglow at
longer wavelengths. This selection is therefore independent of any
{\it a priori\,} information on the properties of the hosts (e.g.,
redshifts, star-formation rates, luminosities, ...). Furthermore it does not take
into account at which wavelength the position of the GRB was
determined, i.e., whether the afterglow was optically bright or
whether it was only seen in the radio and/or the X-rays (i.e.,
``dark'' burst).  
This led to a sample of 15 objects, to which we
added the host of the burst detected on February 22$^{\rm nd}$, 2001
(i.e., GRB\,010222). This host galaxy has been claimed to be
associated with a SCUBA/MAMBO ultra-luminous infrared object
\citep{Frail02}, making it an obvious and quite interesting target for
IR observations.

We note however that our sample is obviously subject to the various
observational cuts that may have biased the identification of GRBs
prior to 2001.  For instance, most of the bursts considered in this
work were detected either with the WCF camera on-board the Beppo-SAX
satellite or with the BATSE instrument on-board the {\it Comptom
  Gamma-Ray Observatory}. As revealed by more recent high energy
missions like {\it HETE-2\,} and {\it Swift\,}, this may have imposed
a first sub-selection over the whole population of GRBs depending on
their intrinsic luminosity and the hardness of their spectrum.
Furthermore, the subsequent ground-based follow-ups that led to the
sub-arcsec localization of these GRBs as well as the determination of
their redshift from optical spectroscopy were likely slower than those
currently operated by the new networks of more dedicated telescopes.
This could in principle favor the detection of X-ray and optical
transients relatively brighter than the afterglows accompagnying the
typical bursts now accessible with {\it Swift} \citep{Berger05}, thus
biasing the selection toward host galaxies with low extinction. In
Sect.\,5.4 we will discuss in more detail the implication of our
results taking into account the above mentioned issues.

The observations were performed with the \spi \, {\it Space Telescope}
\citep{Werner04} as part of the IRS Guaranteed Time Observing program
\citep{Houck04a}.  The J2000 coordinates and the redshifts of the
targeted galaxies are reported in Table\,1 along with the name of the
GRBs that led to their selection. We initiated our \spi \, program in
2004 while these gamma-ray bursts all occured before February 2001.
The contribution of the emission from their fading afterglow was
therefore negligible at the time of our observations.

Each object was imaged with the ``InfraRed Array Camera'' (IRAC,
\citealt{Fazio04a}) at 4.5\mic and 8.0\mic as well as with the
``Multi-band Imager and Photometer for \spi'' (MIPS,
\citealt{Rieke04}) at 24\micpa.  The IRAC detectors are characterized
by 256$\times$256 squared pixel arrays with a pixel size of
1.22\arcsec \, leading to a total field of view of
5.2$\times$5.2\,arcmin$^2$.  The Full-Width at Half Maximum (FWHM) of
the Point Spread Function (PSF) varies between 1.5\arcsec \, and
2\arcsec \, accross the different IRAC channels.  The MIPS detector at
24\mic uses a 2.45\arcsec\,pixel size array of 128$\times$128 elements
also resulting in a field of view of 5.2$\times$5.2\,arcmin$^2$. The
image at this wavelength is characterized by a PSF with a FWHM of
$\sim$6\arcsec.

Each observation was performed using a sequence of several frames
slightly dithered with respect to the position of the GRB host.  Three
frames of 100\,s were obtained at 4.5\mic and 8\mic giving a total
integration time of 300\,s per source and per band.  At 24\micpa, 14
frames of 30\,s led to a total exposure time of 420\,s per source.
The data were reduced with standard procedures (i.e., dark-current
subtraction, cosmic ray removal, non-linearity correction,
flat-fielding and mosaiking) using the pipeline of \spi\, Science
Center\footnote{see http://ssc.spitzer.caltech.edu/postbcd/}.

The absolute pointing accuracy of the \spi \, satellite is better than
$\sim$\,1\arcsec. The 1$\sigma$ relative astrometric uncertainty
is less than $\sim$\,0.3\arcsec \, in the IRAC and MIPS data.

\subsection{IRAC and MIPS Photometry}

With the exception of the GRB\,980425 host galaxy lying at
$z$\,=\,0.0085 \citep{Galama98,Tinney98}, the other GRB hosts of our
sample are all located at cosmological distances ($z$\,$\geq$\,0.84,
see Table\,1). Therefore they are not spatially resolved in the \spi
\, images and their fluxes can be estimated using small circular
aperture photometry.  In the IRAC images, counts were measured over a
circled area with a radius of 3\,pixels (e.g., 3.6\arcsec) centered at
the position of each target. These counts were translated into flux
densities using the conversion factor prescribed in the {\it Spitzer
  Observing Manual\,}\footnote{An electronic version of the {\it
    Spitzer Observing Manual\,} is available at
  http://ssc.spitzer.caltech.edu/documents/SOM/} and a slight
correction was finally applied to account for the extended size of the
PSF. Sensitivity limits of $\sim$3.5\muJy and $\sim$\,20\muJy
(3$\sigma$) were estimated at 4.5\mic and 8.0\mic respectively, based
on the dispersion of the flux measurements obtained over blank sky
regions within the same aperture as the one used for the photometry of
the objects.

Evidence for non-negligible blending was found at 24\micpa, which is
mostly due to the larger size of the PSF and the higher level of
extragalactic confusion in the MIPS images than in the IRAC data. To
ensure accurate results, source extraction and photometry were
therefore performed using the PSF fitting technique of the DAOPHOT
software \citep{Stetson87}. We constructed an empirical point spread
function from the brightest point sources found in our 24\mic
data. This PSF was accordingly scaled to provide the best match to the
object detected at the position of the GRB host, thus leading to a
direct estimate of its total flux at 24\micpa.  As a sanity check, the
same procedure was also performed using a 24\mic theoretical PSF
simulated and provided by the \spi \, Science Center.  Within the
uncertainties, the photometric measurements that we obtained in this
case are consistent with those derived using the empirical
point-spread function.  A 3$\sigma$ sensitivity limit of
$\sim$\,85\muJy was estimated using aperture photometry.

Regarding the nearby GRB\,980425 host galaxy, a diffuse extended
infrared emission was detected up to a distance of 20\arcsec \, to
30\arcsec \, from the center of the object.  The sky background level
was therefore estimated within an annulus defined by an inner radius
of 35\arcsec \, and a width of 10\arcsec.  At 4.5\mic the total flux
of the galaxy was determined using a 23\arcsec-radius aperture. This
allowed us to recover most of the extended emission of the object
while avoiding the contamination from other field sources located
close to the host.  At 8\mic and 24\micpa, a 35\arcsec-radius aperture
was found to provide a very good estimate of the total flux of the
galaxy. At these wavelengths the source density in the field is
smaller than observed at 4.5\micpa, and no other contaminant object
was detected within this large aperture.

Our flux measurements and upper limits are given in Table\,1.  The
absolute photometric uncertainties in the IRAC and the MIPS data are
respectively less than 5\% and 10\%.

\section{Results}

\subsection{Detection rate of the GRB host galaxies in the IRAC and MIPS images}

The 4.5/8.0\mic IRAC and 24\mic MIPS images of the GRB host galaxies
for which a detection in at least one mid-IR band was obtained (see
also Table 1) are presented as postage stamps in Figures\,1 to 6.  To
facilitate the identification of the fields of view, they are
displayed along with optical images publicly available in the
literature. These optical data were obtained with the STIS or the WFPC
cameras on-board the {\it Hubble Space Telescope\,} (HST) as part of
various observing programs led by Fruchter et al.  (HST
Proposals~7966/8189), Holland et al. (HST Prop.:~8640) and Kulkarni et
al. (HST Prop.:~8867).  Most of the reduced images are taken from the
``Survey of the Host Galaxies of Gamma-Ray
Bursts''\footnote{http://www.ifa.au.dk/$\sim$hst/grb\_hosts/intro.html}
\citep{Holland00a}, with the exception of the data for the hosts of
GRB\,970828 and GRB\,010222. For the latter, reduced products were
provided by the ``Multimission Archive at the Space Telescope Science
Institute'' (MAST)\footnote{http://archive.stsci.edu}.  Most of these
HST images are displayed with higher spatial resolution in the
Figure\,2 of \citet{Bloom02a}.

The nearby host of GRB\,980425 is clearly detected in the three bands
that we covered with IRAC and MIPS.  However, we note that most of the
cosmological GRB hosts from our sample {\it are not detected\,} with
\spi. Among the 15 high-redshift targeted sources, 5 are brighter than
3$\sigma$ at 4.5\micpa, while only 1 and 3 objects are detected at
8\mic and 24\mic respectively.  At $z$\,$\sim$\,1 a typical L$_{*}$
galaxy is most often an intermediate-mass IR-luminous spiral
\citep[e.g.,][]{Zheng04,Hammer05,LeFloch05,Melbourne05}, with an SED
leading to observed flux densities of $f_{\rm 4.5\mu
  m}$\,$\sim$\,10\muJypa, $f_{\rm 8\mu m}$\,$\sim$\,8\muJy and $f_{\rm
  24\mu m}$\,$\sim$\,90\muJypa. Given the 3$\sigma$ sensitivity limits
of our IRAC and MIPS data (see Sect.\,2.2), these non-detections
consequently have strong implications on the nature of the GRB hosts
relative to typical field star-forming galaxies.  In Sect.\,4 we will
describe on a case-by-case basis the constraints that can be derived
from these results regarding the spectral energy distributions (SEDs)
of each galaxy. We will discuss their global implications for our
general understanding of the GRB hosts in Sect.\,5.

For each of the three {\it Spitzer\,} bands, the non-detected sources
were also stacked together in the attempt of infering deeper
constraints on the average mid-IR fluxes of these objects considered
as a whole population.  However the small number of stacked images
improved the original depth of our data by only a factor of
$\sim$\,1.5--3 depending on the wavelength, and no signal was detected
above the resulting 3$\sigma$ levels. We note that even though we
performed the stacking test, this approach should ideally be applied
only for sources located in a thin redshift slice, so that the
constraint on the flux measured in the stacking can be converted into
a more physical quantity associated with these objects. As a result,
the broad range of high redshifts covered by our targets (see
Table\,1) prevents a robust interpretation of our lack of detection in
the final stacked GRB host images.  It does suggest though that
considerably deeper observations will be required in order to increase
the rate of GRB host detections with \spi.

\subsection{Infrared luminosities}

In the local Universe, correlations between the mid-IR luminosity and
the 8-1000\mic integrated emission of galaxies have been observed not
only for normal and quiescently star-forming objects
\citep{Dale01,Roussel01} but also for more actively starbursting
sources \citep{Chary01}.  Such correlations likely hold also in the
distant Universe, as the mid-IR/radio relationship resulting from the
mid-IR/far-IR and far-IR/radio local correlations is still observed at
high redshifts \citep{Gruppioni03,Appleton04}.  Therefore the mid-IR
\spi \, data can potentially be used to infer some constraints on the
total infrared luminosity of distant galaxies.

In this extrapolation, the errors are largely dominated by the
uncertainties on the shape of the underlying SED from 8\mic to
1000\micpa. For example, the predictions from the libraries of
starburst-dominated SEDs proposed by \citet{Chary01}, \citet{Dale01}
and \citet{Lagache03} appear to be consistent within only 0.3\,dex up
to $z$\,$\sim$\,2, but this dispersion can be significantly larger
assuming other SEDs \citep[e.g.,][]{Dale05}. At $z$\,$\gtapp$\,1.5 in
fact, our 24\mic data are only sensitive to the brightest galaxies
(i.e., ULIRGs), which harbor a large diversity of mid-IR properties
\citep[e.g.,][, Armus et al. 2006 in prep.]{Armus04}.  The prediction
of their total IR luminosity based on a single mid-infrared flux
measurement can be uncertain by a factor of~$\gtapp$\,5 for a given
object.

To constrain the bolometric luminosity of the GRB host galaxies, we
first converted the 24\mic flux density (or upper limit) measured for
each object with a known spectroscopic redshift to a monochromatic
luminosity at the corresponding rest-frame wavelength
24\micpa/(1+$z$). Following the approach presented by
\citet{LeFloch05}, this estimate was then translated into a total IR
luminosity using the three libraries mentioned in the previous
paragraph.  We should emphasize that in these collections of IR
spectra, a given monochromatic luminosity at a given wavelength
corresponds to a single ``total IR'' luminosity, thus leading to a
unique determination of this quantity for each 24\mic flux measured in
our sample.  The estimates derived from the three libraries were
therefore averaged for each object and their dispersion was used to
quantify the associated uncertainty.  We consider our results to be
accurate within a factor of $\sim$\,2--3 up to $z$\,$\sim$\,1.5 and
within a factor of $\sim$\,3--5 at higher redshifts. We also keep in
mind that such uncertainties apply on a case-by-case basis, and they
are obviously smaller when addressing the average IR luminosity of a
sample of galaxies.

 These estimates are reported in Table\,1 along with the equivalent
 star formation rates (SFR) derived from these infrared luminosities
 using the calibration proposed by \citet{Kennicutt98}. These
 conversions assume that the whole IR emission detected at 24\mic
 originates from star-forming activity.  While the host of GRB\,980425
 is only a modest infrared emitter (L$_{\rm
   IR}$\,=\,2\,$\times$\,10$^{9}$\,L$_\odot$), the three
 24\micpa-detected high-redshift sources (i.e., the hosts of
 GRB\,970828, GRB\,980613 and GRB\,990705) are characterized by a
 total infrared luminosity \Lir \, in the range of
 10$^{11}$\,L$_\odot$\,$\leq$\,L$_{\rm
   IR}$\,$\leq$\,10$^{12}$\,L$_\odot$, bringing them to the class of
 the so-called Luminous InfraRed Galaxies \citep[LIRGs,][]{Sanders96}.
 However, there is no detection of a host with luminosity larger than
 10$^{12}$\,L$_\odot$ in the range of the ULIRGs (Ultra-Luminous
 Infrared Galaxies: 10$^{12}$\,L$_\odot$\,$\leq$\,L$_{\rm
   IR}$\,$\leq$\,10$^{13}$\,L$_\odot$) and the HyLIRGs (Hyper-Luminous
 Infrared Galaxies: L$_{\rm IR}$\,$\geq$\,10$^{13}$\,L$_\odot$).  In
 the case of the non-detections, the measured 3$\sigma$ sensitivities
 were used to derive an upper limit on the total IR luminosity when a
 confirmed redshift was available.  Most of these constraints are also
 consistent with infrared luminosities lower than
 10$^{12}$\,L$_\odot$, even though we can not definitely rule out
 having a few GRB hosts with ULIRG-type IR luminosities given the
 uncertainties affecting our estimates.

One might note a few potential caveats that could affect these total
IR luminosity estimates.  First, we can not completely exclude the
presence of a dust-embedded active galactic nucleus (AGN) lurking in
these GRB host galaxies and dominating their mid-IR emission. Because
the IR SED of AGNs is generally much flatter than the
starburst-dominated SEDs that we assumed in the conversion of the
24\mic flux density \citep[see][]{Weedman05}, our determination of the
total IR luminosity could be over-estimated by a factor of 5 to 10 in
these cases.  However, we consider this possibility unlikely as no
typical AGN signature has ever been reported from the optical, X-ray
and radio properties of these objects.

Furthermore, we have implicitly assumed that the MIPS 24\mic
detections purely originate from star-forming activity in the hosts
and we have neglected a possible contribution from a transient
emission due to the effect of the GRB on its close environment. Based
on a detailed modeling of the heating effect of GRBs and their
afterflows on their surrounding region, \citet{Venemans01} have argued
that a reprocessed dust mid-IR emission could in principle be easily
detected by \spi \, several years after a burst occuring in a dusty
star-forming galaxy. In our analysis of the IR spectral energy
distributions of the GRB hosts, presented in Sections\,4 \&~5, we do
not consider this likelihood. We will argue though that the
contribution of such a GRB dust emission is unlikely to be significant
in our sample.

\subsection{Spectral energy distributions}

The GRB host galaxies of our sample have already been extensively
observed at optical and near-infrared wavelengths, and some of them
have also been targeted by submillimeter and radio observations.
Consequently, our \spi \, data can be used in a multi-wavelength
context to derive some constraints on the global spectral energy
distribution and the nature of these sources.

We combined our mid-infrared photometry with other broad-band imaging
data retrieved from the literature (see the caption of Fig.\,7 for
references).  Optical and near-infrared magnitudes were first
corrected from the foreground Galactic extinction using the DIRBE/IRAS
dust maps of \citet{Schlegel98} and assuming the $R_V$\,=\,3.1
extinction curve of \citet{Cardelli89}. Depending on the quoted
magnitude reference, they were then converted into fluxes using the
zero points from the Vega or the AB systems. These fluxes as well as
the fluxes or the upper limits gathered at the other wavelengths were
finally converted into rest-frame monochromatic luminosities to
provide constraints on the GRB host SEDs from the optical up to the
mid-IR or the submillimeter/radio wavelength range.

Our results are illustrated in Figure\,7. With the exception of the
GRB\,990510 host galaxy that we did not detect with \spi \, and which
is also optically very faint
\citep[$V$\,$\sim$\,28\,mag,][]{Fruchter00c}, all GRB hosts with a
confirmed spectroscopic redshift in our sample are shown.  Given the
very low rate of detections at long wavelengths (i.e.,
$\lambda$\,$\gtapp$\,3\micpa), we do not perform a fit to the current
data and we refer the reader to e.g., \citet{Sokolov01},
\citet{Christensen04} and \citet{Chary02} for a statistical stellar
population synthesis of the GRB host optical and near-IR
properties. Rather, we consider as a comparison the spectral energy
distribution of the prototypical objects NGC\,253, M\,82 and Mrk\,231.
NGC\,253 and M\,82 are moderately-active star-forming spirals
characterized by IR luminosities L$_{\rm
  IR}$\,$\sim$\,1.4$\times$10$^{10}$\,L$_\odot$ and L$_{\rm
  IR}$\,$\sim$\,3.7$\times$10$^{10}$\,L$_\odot$ respectively
\citep{ForsterSchreiber03}. Their mid-IR SEDs were derived from the
ISOCAM-CVF spectra of \citet{ForsterSchreiber03}, which were further
extrapolated to the UV/optical and the far-IR/radio domains using the
modelling provided by \citet{Silva98}.  Mrk\,231, on the other hand,
is a warm ULIRG with an IR luminosity L$_{\rm
  IR}$\,$\sim$\,4$\times$10$^{12}$\,L$_\odot$ powered both by a
luminous AGN and a violent starburst \citep{Farrah03}. Its SED was
derived by combining the 5--35\mic spectrum recently obtained by the
IRS on-board \spi \, \citep[; Armus et al., in prep.]{Weedman05} with
far-IR and radio data published in the literature
\citep[e.g.,][]{Ivison04}.  In addition to these three sources, we
also displayed the template of a starburst-dominated ULIRG with
L$_{\rm IR}$\,=\,4\,$\times$\,10$^{12}$\,\Lsol \, as well as the SED
of a cold LIRG with L$_{\rm IR}$\,=\,10$^{11}$\,\Lsol.  These two
spectral energy distributions were taken from the IR galaxy libraries
derived by \citet{Chary01} and \citet{Lagache03} respectively.

All these SEDs were chosen to be globally representative of the
expected emission from starburst galaxies and IR-luminous sources of
the local Universe.  They are displayed in Figure\,7 with the y-axis
in units of W\,Hz$^{-1}$ in order to provide a simple and direct
qualitative comparison with the observed properties of the GRB hosts
as a function of wavelength. As a result they should only be viewed in
this context.  Since no fitting to the detections and upper limits was
attempted, some of these SEDs were actually re-scaled to match the
luminosity of the GRB hosts at certain wavelengths. This normalization
was done in an {\it ad-hoc\,} manner, either at the shortest detected
wavelengths (e.g., $B$-band), in the rest-frame near-IR (e.g.,
$K$-band or IRAC~4.5\mic channel), at the MIPS~24\mic observed
wavelength, or even in the radio as in the case of the GRB\,980703
host.  It better reveals how the global SEDs of the GRB-selected
galaxies deviate from the other templates considered in the figure,
and it also shows the large diversity of properties characterizing the
GRB host population.
This will be more
thoroughly discussed on a galaxy-by-galaxy basis in Sect.\,4.

\section{The GRB hosts in a multiwavelength context}

Following the results derived in Sect.\,3.3 we analyze hereafter the
multi-wavelength properties of the GRB host galaxies using the
spectral energy distributions displayed in Fig.\,7.  We first discuss
on a source-by-source basis the objects with a known spectroscopic
redshift and ordered by increasing distance from Earth, and we briefly
mention the properties of the remaining sources at the end of the
section.

\subsection{The host galaxy of the GRB\,980425}

The association of the GRB\,980425 with the SN\,1998bw initially
proposed by \citet{Galama98} led to the identification of a nearby
sub-luminous blue galaxy located at z=0.0085.  Classified as an Sbc
type in the Hubble sequence \citep{Fynbo00}, it is currently the only
GRB-selected object known in the local Universe.  The supernova
SN\,1998bw was observed in one of its spiral arms at a distance of
$\sim$\,900\,pc from a rather bright HII region.
 
As shown in Fig.\,2 this galaxy is detected with a high signal to
noise in the three \spi \, bands.  Because of its proximity it is also
well resolved and a diffuse emission extending along the major axis is
clearly observed. Nonetheless the most striking result is the
detection of a bright mid-IR point source located very close to the
region where the GRB occured. It is the brightest point source
detected at 4.5\mic and its contribution to the total emission of the
galaxy is steeply rising with wavelength. With a flux density of
220\muJy and 1815\muJy at 4.5\mic and 8\micpa, it represents
respectively 7\% and 15\% of the monochromatic luminosity of the whole
galaxy measured in the IRAC channels.  At 24\mic its flux reaches
$\sim$21\,mJy, implying that more than $\sim$75\% of the energy
radiated by the galaxy at the MIPS wavelength arises from this region.

This luminous point source represents one the reddest objects so far
identified in the nearby Universe. With a 24\mic monochromatic
luminosity $L_{\rm 24\mu m}$\,$\sim$\,1.1\,$\times$\,10$^8$\,\Lsol, it
is also much more luminous than W49 which is the brightest HII region
within the Milky Way \citep{Harper71}.  As a result, its discovery in
a GRB-selected galaxy raises obviously the question of a physical link
with the hypernova.  In spite of the large PSF characterizing the \spi
\, data, the brightness of this object allowed us to estimate its
position along the spiral arm of the galaxy with an accuracy better
than 0.4\arcsec. We believe that it is unlikely to be related to the
close environment of the GRB but it rather coincides with the bright
HII region located $\sim$5\arcsec \, (i.e., $\sim$900\,pc) in the
North-West direction. It could be due, for instance, to a dense super
star cluster deeply embedded in dust. This source is to be observed
very soon with the Infrared Spectrograph of \spi \, to explore in more
detail its nature and its mid-IR properties, and the results will be
reported in a forthcoming paper.

Should this object be indeed related to intense dusty star formation,
our mid-IR observations as well as the optical view of the galaxy
reveal that the GRB did not occur in the most active site of star
formation within its host.

\subsection{The host galaxy of GRB\,970508}

This object is a blue compact dwarf galaxy \citep{Fruchter00b} located
at $z$\,=\,0.835 and showing at optical wavelengths a quiescent
activity of star formation \citep{Bloom98a}.  Its 4000--8000\AA \,
spectrum reveals a blue continuum with an [OII] emission line
corresponding to a star formation rate $SFR_{\rm \,
  [OII]}$\,$\sim$\,1\,M$_\odot$\,yr$^{-1}$ \citep[not corrected for
  dust extinction,][]{Bloom98a}.  It is not detected with IRAC nor
with MIPS. Given its moderate redshift and the depth of our data, this
non-detection with \spi \, and the flux density measured in the
$K$-band exclude the presence of a massive underlying stellar
population in this object. It also argues against a significant
contribution from dust-obscured star-forming activity. As a result the
total star formation rate is unlikely to be much larger than the $SFR$
derived from the optical. Based on the sensitivity of our 24\mic data
we infer an upper limit of 10$^{11}$\,\Lsol \, for its total IR
luminosity.

Note that the possible association of the GRB\,970508 host with an
ultra-luminous infrared galaxy proposed by \citet{Hanlon00} based on
ISOPHOT observations is clearly ruled out by the \spi \, data.

\subsection{The host galaxy of GRB\,990705}
\label{sec:990705}

The GRB\,990705 occured within a large and optically-luminous
star-forming spiral galaxy observed face-on at $z$\,=\,0.8424
\citep{LeFloch02a}. The absolute magnitude of this GRB host
corresponds to a 2\,L$_\star$ galaxy at $z$\,$\sim$\,1.  While GRBs
preferentially occur within sub-luminous and young sources, the
luminosity and the morphology of this object indicates therefore that
some bursts can also take place within luminous and more evolved
systems.

The galaxy is clearly detected at 4.5\mic with IRAC and 24\mic with
MIPS.  Based on the MIPS detection we estimate a total infrared
luminosity L$_{\rm IR}$\,=\,1.8$^{+2.1}_{-0.6}\times$10$^{11}$\Lsol \,
(see Table\,1).  Hence, this source belongs to the category of the
Luminous Infrared Galaxies.  Assuming that the totality of the
infrared emission is powered by star formation, its luminosity
corresponds to a star formation rate $SFR_{\rm
  \,ir}$\,$\sim$\,32$^{+37}_{-11}$\,\Msol\,yr$^{-1}$, which is
somewhat larger than the SFR derived from the observed UV continuum
\citep[$SFR_{\rm \, uv}$\,$\sim$\,5--8\,M$_\odot$,][]{LeFloch02a}.
These characteristics and the observed morphology of the source are
actually very similar to those of the IR-luminous spirals that were
detected in the $ISO$ and {\it Spitzer} deep surveys
\citep[e.g.,][]{Flores99,Zheng04,Bell05,Melbourne05} and that dominate
the star formation at $z$\,$\sim$\,1
\citep[e.g.,][]{Chary01,LeFloch05}.  Contrary to many other
GRB-selected galaxies, the host of GRB\,990705 is therefore a very
good representative of the sources responsible for the bulk of the
star-forming activity in the Universe at such redshifts.

As shown in Figure\,7, the lack of data from the $I$, $J$, $H$ or $K$
bands results in a rather poor sampling of the SED in the rest-frame
optical and near-IR. This prevents us from infering the exact nature
of the emission that dominates the luminosity of this galaxy at these
wavelengths. However the non-detection at 8\mic provides an additional
strong constraint that firmly excludes the presence of hot dust
dominating at short mid-IR wavelengths. As illustrated by the
comparison with the local templates, the IRAC detection at 4.5\mic
thus argues for an evolved and massive underlying stellar population
dominating the near-IR emission.

\subsection{The host galaxy of GRB\,970828}
\label{sec:970828}

The host of GRB\,970828 appears as an early-stage interacting system
at $z$\,=\,0.96, with 3 components (refered to as galaxies ``A'',
``B'' and ``C'') located over a $\sim$30\,kpc region
\citep{Djorgovski01a,Bloom01c}.  The galaxy where the burst is
believed to have occured (component ``galaxy B'') is faint at optical
wavelengths ($R$\,=\,25.1) but it is one of the reddest GRB hosts
($R-K$\,=\,3.6, \citealt{Djorgovski01a}) in the near-infrared sample
studied by \citet{LeFloch03}.

There is a clear detection of this merging system at 4.5\mic and
24\micpa. In the IRAC image it appears as a point source rougly
centered between the two galaxies ``A'' and ``B''. These two
components are only separated by $\sim$\,1.9\arcsec \, on the sky
\citep{Bloom01c}. They can not be resolved with \spi \, and it is not
possible to unambiguously determine whether the infrared emission
originates from only one or both of these objects.

Based on the MIPS detection we estimate a total infrared luminosity
\Lir\,=\,1.4$^{+2.5}_{-0.8}\times$10$^{11}$\Lsol.  This merging system
is therefore a Luminous Infrared Galaxy like the host of GRB\,990705,
characterized by a dust-obscured star formation rate $SFR_{\rm \,
  ir}$\,$\sim$\,24$^{+43}_{-14}$\,\Msol yr$^{-1}$.  This $SFR$\,
estimate is much larger than the one derived from the UV continuum or
from the flux of the [OII] emission line detected in the Keck spectra
for the two galaxies ``A'' and ``B'' \citep[$SFR_{\rm \,
    uv}$\,$\sim$\,$SFR_{\rm \,
    [OII]}$\,$\sim$\,0.5--1\,\Msol\,yr$^{-1}$,][]{Djorgovski01a}.  As
shown in Fig.\,7 the spectral energy distribution is steeply rising
from the optical to the mid-infrared, which suggests the presence of a
typical dust-enshrouded young starburst with no underlying evolved
stellar population.

Interestingly enough the GRB\,970828 is generally refered to as a
typical ``dark burst''.  The optical afterglow of this GRB was not
detected despite a deep and prompt search down to
$R$\,$\sim$\,24.5\,mag and despite an accurate localization of the
event thanks to the detection of its radio transient
counterpart. Given the moderately-high redshift of its host galaxy,
\citet{Djorgovski01a} thus suggested that its optical emission was
likely suppressed by an intervening cloud of material within the host,
invoking the dust extinction as one possible explanation for the
origin of at least a fraction of these ``dark GRBs''. Our infrared
detection with \spi \, provides obviously a strong support for this
hypothesis. In luminous-infrared galaxies, most of the UV radiation
emitted by young and massive stars are indeed absorbed by dust and
re-radiated at longer wavelengths.

\subsection{The host galaxy of GRB\,980703}

This object is one of few GRB hosts characterized by an unambiguous
detection at radio wavelengths \citep{Berger01b}. Observations
performed at the VLA after the fading of the GRB radio counterpart
revealed at the location of the burst a persistent emission of
68.0\,$\pm$\,6.6\muJy at 1.43\,GHz.  Assuming the far-infrared/radio
correlation observed in local starburst galaxies \citep{Condon92},
\citet{Berger01b} have thus argued that the host of GRB\,980703 is an
ultra-luminous infrared galaxy (i.e., \Lir\,$\sim$\,10$^{12}$\,\Lsol)
characterized by an $SFR$\, of several hundreds of \Msol\,yr$^{-1}$.

At the redshift of this object ($z$\,=\,0.97, \citealt{Djorgovski98}),
ULIRGs are very easily detected at the mid-IR \spi \, wavelengths
(i.e., IRAC + MIPS 24\micpa).  Strikingly though, the GRB\,980703 host
galaxy is not detected in our data, except in the IRAC-4.5\mic
channel.  As shown in Fig.\,7 its spectral energy distribution from
the mid-IR to the radio is therefore not consistent with typical
starburst-dominated infrared-luminous galaxies, and our non-detection
at 24\mic argues for a very modest star-formation rate ($SFR_{\rm \,
  ir}$\,$\ltapp$\,24\,\Msol\,yr$^{-1}$) when compared to the $SFR$
claimed by \citet{Berger01b}.

The excess of radio emission relative to the mid-IR flux can be
quantified by the $q_{24}$ parameter defined as $q_{24}=log_{10}
(S_{24\mu m}/S_{20cm})$, where $S_{\lambda}$ is the observed
monochromatic flux density at the considered wavelength
\citep{Appleton04}. Taking into account the sensitivity of our 24\mic
data, we derive an upper limit of $q_{24}=0.09$ for the host of
GRB\,980703, which is significantly lower than the average value
$q_{24} \sim 1$ observed for starburst galaxies \citep{Appleton04}.

Low $q_{24}$ measurements (i.e., $q_{24}$\,$\ltapp$\,0) are usually
interpreted as the signature of radio-loud objects characterized by an
AGN-dominated SED \citep{Higdon05}.  Would the presence of an active
galactic nucleus explain this radio detection in the host of
GRB\,980703\,?  There is no evidence favoring this hypothesis in the
optical spectrum of the host \citep{Djorgovski98}, and no temporal
variability in the radio emission has been reported by
\citet{Berger01b}.  However we note that the slope of the radio
continuum (spectral index $\beta$\,$\sim$\,0.3, where $F_{\nu} \propto
\nu ^{-\beta}$) is significantly flatter than the spectrum
characterizing the majority of the starburst ``\muJy galaxies'' (i.e.,
$S_{20cm}$\,$\ltapp$\,100\muJy) selected at radio wavelengths
($\beta$\,$\sim$\,0.8, \citealt{Richards00}).  As a result, the radio
spectrum in itself would be more easily explained by the presence of
an AGN than supernova remnants in starbursting regions. In fact, the
existing data do not allow us to unambiguously disentangle between
these two potential contributions, and this GRB host could still be
characterized by a rare type of IR-luminous SED yet leading to a faint
emission at 24\micpa.  Deep X-ray observations and/or far-IR MIPS
imaging at 70\mic and 160\mic should provide better constraints on the
nature of this object and its spectral energy distribution.

\subsection{The host galaxy of GRB\,980613}

The host of GRB\,980613 is another merging system characterized by a
very complex environment with up to 9~galaxy fragments interacting
with each other at $z$\,=\,1.10 \citep{Chary02,Hjorth02,Djorgovski03}.
These components show a moderate activity of star formation in the
optical (i.e., $SFR_{\rm \, opt.}$\,$\ltapp$\,5\,\Msol\,yr$^{-1}$) but
some of them display very red $R-K$ colors.

The system is clearly apparent in our \spi \, data.  In addition to
the detections at 4.5\mic and 24\mic it is actually the only host
galaxy of our sample also detected at 8\micpa.  The flux estimated at
24\mic corresponds to a total infrared luminosity
\Lir\,$=$\,5$^{+9}_{-3}\times$10$^{11}$\,\Lsol, leading to an
IR-equivalent star formation rate $SFR_{\rm \,
  ir}$\,$\sim$\,87$^{+156}_{-52}$\,\Msol\,yr$^{-1}$.

However it should be noted that the \spi \, detection does not
coincide with the component of the interaction where the GRB was
observed (component ``H'', see \citealt{Hjorth02}). It rather
corresponds to the very red fragments denoted ``C'' and ``D'' by
\citet{Chary02} and \citet{Djorgovski03}.  These two components are
not spatially resolved by \spi \, but they are located more than
2.5\arcsec \, away from component ``H'' and they can be distinguished
from the latter.  We infer that the GRB\,980613 did not occur in the
region harboring the most intense star-forming activity of the system,
as was already noted by \citet{Hjorth02} based on deep optical HST
data.

The spectral energy distribution of the component detected with \spi
\, in this interaction presents a striking contrast with the SEDs of
the other GRB hosts observed in our data (see Figure\,7). The fluxes
measured at 4.5\mic and 8\mic are particularly bright given the
redshift of the host galaxy, and the 8\mic detection reveals a clear
inflexion of the SED in the rest-frame near-infrared.  This suggests
not only a significant hot dust emission dominating the SED redward
$\sim$\,2\mic but also the contribution of an evolved and massive
underlying stellar population likely dominating the optical
wavelengths.

\subsection{The host galaxy of GRB\,990506}

The GRB\,990506 is another typical example of a ``dark burst''
undetected in the optical despite deep and prompt imaging after the
gamma-ray explosion.  Its accurate localization was obtained based on
the detection of its radio afterglow \citep{Taylor00} and subsequent
follow-ups revealed a host galaxy located at $z$\,=\,1.31
\citep{Bloom03}, characterized by a very compact morphology
\citep{Holland00d} and a red $R-K$ color
\citep[$R-K$\,$\sim$\,4,][]{LeFloch03}.  Its [OII] emission line
indicates a dust-uncorrected star formation rate $SFR_{\rm \,
  [OII]}$\,=\,13\,\Msol\,yr$^{-1}$ \citep{Bloom03}, which is
substantially higher than the median $SFR$ \, characterizing the
global population of the GRB host galaxies at optical wavelenghts.

It is marginally detected at the 2$\sigma$ level at 4.5\micpa, but it
is not seen in the 8\mic and 24\mic images. As already suggested by
\citet{Barnard03} who did not detect this object with SCUBA, this case
reveals that ``dark GRBs'' are not systematically associated with
dust-enshrouded massive star-forming activity.

\subsection{The host galaxy of GRB\,010222}

The GRB\,010222 has been widely refered as the prototypical burst
associated with violent starburst activity at high redshift, bringing
also further support for the death of young and massive stars as the
origin of long GRBs.  A persistent source with an average flux density
of 3.74\,$\pm$\,0.53\,mJy at 850\mic and 1.05\,$\pm$\,0.22\,mJy at
1.2\,mm was indeed observed with SCUBA and MAMBO at the location of
its afterglow \citep{Frail02}.  At the redshift of the burst
\citep[$z$\,=\,1.48,][]{Jha01}, the reported flux densities correspond
to an infrared luminosity \Lir\,$\sim$\,4$\times$10$^{12}$\,\Lsol \,
assuming typical far-infrared galaxy SEDs. Therefore \citet{Frail02}
argued that the GRB\,010222 host is a dusty ultra-luminous infrared
galaxy experiencing a very intense episode of star formation
($SFR$\,$\sim$\,600\,\Msol\,yr$^{-1}$).

However, this object is not detected in our data (see Figure\,8),
which obviously raises some doubt regarding its association with the
SCUBA/MAMBO source.  The typical SCUBA and MAMBO galaxies have easily
been detected with \spi \,
\citep{Charmandaris04b,Egami04a,Frayer04,Ivison04} while they are
located on average at higher redshifts than the host of GRB\,010222
\citep[e.g.,][]{Chapman03}. Furthermore, this GRB host galaxy is a
very faint blue object that contrasts with the typical properties
characterizing the optical counterparts of the SCUBA sources
\citep{Smail04}. Considering that it is not detected with more than
3$\sigma$ in any of the Spitzer mid-IR bands nor at radio wavelenfgths
\citep{Berger03}, the claim for an association between the GRB host
and the SCUBA/MAMBO source thus appears questionable.  In fact
\citet{Frail02} mention the presence of 4 other redder galaxies
detected in the $K$-band and located within the 15\arcsec-diameter
beam of SCUBA centered at the position of the GRB host.  Three of
these galaxies {\it are\,} detected with IRAC and with MIPS, which
could suggest that the SCUBA/MAMBO detection is likely associated with
one or several of these other sources instead.

Strictly speaking though, the non-detection of the host at 24\mic is
not sufficient to completely rule out the possible association between
the GRB\,010222 and a ULIRG. The SED of ultra-luminous infrared
galaxies can be characterized by a strong silicate absorption at
9.7\mic rest-frame \citep{Spoon04,Armus04}. At the distance of the
host ($z$\,=\,1.48), this feature would be redshifted in the 24\mic
band, which could explain the non-detection by MIPS at this wavelength
\citep[see][ for detailed discussion on this issue]{Kasliwal05}. In
this case however, the non-detection at 4.5\mic would still remain
very puzzling.

\subsection{The host galaxy of GRB\,990123}

The high spatial resolution image of this galaxy obtained with the HST
revealed a strongly-interacting system with a complex morphology, and
the optical transient of GRB\,990123 was actually observed very near a
star-forming region associated with one of its merging components
\citep{Bloom99,Fruchter99a,Holland99}.  Located at $z$\,=\,1.6
\citep{Kulkarni99}, the GRB host has a luminosity of
$\sim$\,0.5\,$L_\star$ in the optical and its observed UV continuum
argues for a small amount of star formation
\citep[$SFR$\,$\sim$\,4\,\Msol\,yr$^{-1}$,][]{Bloom99}.

Our IRAC and MIPS observations did not lead to any detection in the
infrared, which indicates that this sub-luminous galaxy is a low-mass
quiescent starburst with no significant contribution from
dust-enshrouded star-forming activity.

\subsection{The host galaxy of GRB\,990510}
\label{sec:990510}

This host galaxy is a very faint object
\citep[$V$\,$\sim$\,28\,mag,][]{Fruchter00c} at $z$\,=\,1.62
\citep{Vreeswijk01} with an absolute $B$-band magnitude
$M_B$\,$\sim$\,--17.20\,mag \citep{LeFloch03}. We did not detect it with
IRAC nor with MIPS.  This suggests that the faintness of this source
observed in the optical does not originate from a high amount of dust
extinction but rather points to an intrinsically low-mass and young
object with negligible amount of star formation.

\subsection{The host galaxy of GRB\,971214}

The host of GRB\,971214 is one of the most distant objects that have
been spectroscopically identified based on the optical transient of a
GRB. Located at a redshift $z$\,=\,3.418, it is an $\sim$\,$L_\star$
galaxy with a somewhat irregular morphology and a surface brightness
probably dominated by an exponential profile
\citep{Kulkarni98,Odewahn98}. Neglecting a possible extinction by
dust, its UV continuum and Ly$\alpha$ emission line both argue for a
rather small star formation rate
\citep[$SFR$\,$\sim$\,1--5\,\Msol\,yr$^{-1}$,][]{Kulkarni98}.

We did not detect this galaxy with \spi.  At the redshift of the GRB,
our 24\mic data are only sensitive to Hyper Luminous Infrared Galaxies
(\Lir\,$\geq$\,10$^{13}$\,\Lsol) at the flux limit of our
observations.  Therefore the non-detection by MIPS does not bring a
strong constraint on the infrared properties of the host.  At similar
redshifts however, typical Lyman-Break Galaxies (LBGs) are easily
detected at the short IRAC wavelengths and some of them are also
detected at 24\mic \citep{Barmby04,Huang05}.  Hence, this lack of
detection in our data indicates that the host of GRB\,971214 is likely
not as massive and evolved as the LBGs coexisting at the same
epoch. It is not one of the most actively-starbursting LBGs neither.

\subsection{The host galaxies of GRB\,980326, GRB\,980329,
GRB\,980519 and GRB\,990308} 

The redshifts of these four GRB host galaxies have not been determined
spectroscopically.  Deep observations performed with the HST have
shown that they are very faint in the optical
\citep[$R$\,$\gtapp$\,26\,mag,][]{Jaunsen03,Fruchter01d}, and none of
them is actually detected with {\it Spitzer}. This suggests that their
faint optical emission is probably not due to a large extinction by
dust, but rather indicates galaxies with very low bolometric
luminosities. In favor of this interpretation, we note that the very
faint host galaxy of GRB\,990510 ($V$\,$\sim$28\,mag, $z$\,=\,1.62)
was neither detected in our sample in spite of its redshift easily
accessible for \spi\, (see Sect.\,\ref{sec:990510}).  The possibility
that these objects are at very high redshift can not be excluded
though \citep[see e.g.,][]{Fruchter99}, in which case the
non-detections with IRAC and MIPS would not be surprising.

\subsection{The host galaxy of GRB\,981226}

As in the case of the four sources previously discussed, the
spectroscopic redshift of the GRB\,981226 host has not been
established.  However it is brighter in the optical
\citep[$R$\,$\sim$\,24.5,][]{Frail99,Holland00b,Saracco01a}.  It has
been detected in the $K_s$ band
\citep[$K_s$\,=\,21.1\,$\pm$\,0.2,][]{LeFloch03}, and it is also
clearly apparent in the IRAC 4.5\mic image. Its $R-K$ color makes it
one of the reddest GRB host galaxies that have been studied so far,
and its SED also looks rather steep between the $K_s$ band and the
IRAC 4.5\mic channel.

Our current data set does not allow us to distinguish whether the IRAC
detection and the red colors of this host reveals an old star
population, a hot dust emission, or a combination of both. Invoking
the presence of dust may be attractive since the GRB\,981226 is also
one of the few ``dark bursts'' that did not exhibit any detectable
optical counterpart.  However we consider this explanation unlikely as
this object is not detected at 8.0\mic or 24\micpa. Note that it was
neither detected in the submillimeter \citep[$F_{\rm 850\mu
    m}$\,=\,--2.79\,$\pm$\,1.17\,mJy,][]{Barnard03} and the flux
reported by \citet{Berger03} at radio wavelengths points to less than
a 2$\sigma$ detection ($F_{\rm
  8.46\,GHz}$\,=\,21\,$\pm$\,12\,\muJypa).  The interpretation is
further complicated by the fact that no spectroscopic redshift has
ever been determined for this galaxy.

\section{Discussion}

\subsection{The origin of the infrared emission in the GRB host galaxies}

The IRAC and MIPS instruments on-board \spi \, have recently opened
new exciting perspectives for tackling the evolution of galaxies in
the early Universe. With unprecedented sensitivity at 3.6--8.0\micpa,
the IRAC channels can sample the rest-frame near-IR emission of very
distant sources, while the MIPS imager can detect the hot dust
emission of active starbursts and AGNs up to $z$\,$\sim$\,2--3.  Both
cameras provide therefore new insights to study the evolved stellar
populations and the importance of dust-obscured star formation at high
redshift.

Interpreting the observed mid-IR emission of GRB-selected galaxies can
be however more subtle than analyzing the other mid-IR sources
detected in the field. As mentioned in Sect.\,3.2 \citet{Venemans01}
have argued that optical and UV afterglows can be significantly
absorbed by dusty material surrounding GRBs, and that the reprocessed
IR light can be observed several years after a burst because of the
very slow time-scale variations of the heated dust emission.  The SED
of this dust component would be characterized by a steep increase from
the short wavelengths up to $\sim$\,8--10\mic rest-frame, followed by
a gentle decline in the mid-IR and far-IR \citep[see Fig.\,5 of
][]{Venemans01}.  Up to redshift $z$\,$\sim$\,1--2, it could dominate
the total luminosity of the host galaxy, then questioning the IRAC and
MIPS flux measurements as mass and star formation rate indicators.

However we did not find any obvious evidence for such ``GRB-heated''
dust emission among the \spi-detected GRB hosts. In the resolved
GRB\,980425 host galaxy at $z$\,=\,0.0085 we do not see any signature
of this effect at the location of the hypernova SN\,1998bw, while the
infrared emission detected in the complex environment of the host of
GRB\,980613 ($z$\,=\,1.10) does not coincide with the galaxy fragment
where the burst occured.  At the redshift of the other sources (i.e.,
0.84\,$\leq$\,$z$\,$\leq$1.1), the IRAC-8\mic and MIPS-24\mic bands
constrain the rest-frame 4\micpa/12\mic flux ratios, which appear to
be too red compared to the SED of the GRB dust component predicted by
the models.

The detectability of this GRB dust emission obviously depends on the
efficiency of the absorption of the UV/optical afterglow.  As already
noted by \citet{Venemans01}, it should be preferentially observed
toward the host galaxies of those ``dark bursts'' originating from
dust-enshrouded star-forming regions\footnote{The so-called ``dark
  bursts'' can also be GRBs with intrinsically faint afterglows
  \citep[e.g.,][]{Fox03} or very high redshift bursts with optical
  counterparts suppressed by Ly$\alpha$ absorption
  \citep[e.g.,][]{Lamb00}.}.  Even though three~of our galaxies fall
indeed in this category (i.e., the hosts of GRB\,970828, GRB\,981226
and GRB\,990506), our sample has been mostly built from GRBs
pinpointed with optical afterglows, and our selection could also
induce a bias against the detection of this burst-remnant dust
emission.

As a result, we conclude that the contribution of this component is
negligible in our data.  The IRAC and MIPS detections should truly
reflect the properties of the stellar populations and the global
star-forming activity within the observed host galaxies.

\subsection{A panchromatic view on the nature of the GRB hosts}

A few sources from our sample display clear signatures of evolved
stellar populations and intense starbursting activity. An example is
the host of GRB\,990705 at $z$\,=\,0.84, which harbors active star
formation ($SFR_{\rm \, ir}$\,$\sim$\,32\,\Msol\,yr$^{-1}$, see
Sect.\,\ref{sec:990705}) and appears as a large and massive Sc spiral
galaxy typical of the disk-dominated systems at these redshifts. Other
cases of active starbursts, like the host of GRB\,000418, have
previously been found at radio and submillimeter wavelengths
\citep{Berger03}.

On average though, our MIPS observations combined with the
submillimeter and radio published photometry do not really favor a
population of GRB host galaxies dominated by very active and luminous
dusty starbursts.  With the exception of the host of GRB\,980703 that
we discussed in Sect.\,4.5, our 24\mic non-detections are consistent
with the existing SCUBA and VLA data and they are even more
constraining if one assumes typical IR-luminous and
starburst-dominated SEDs for the host of GRBs (see Figure\,7). Our
measurements argue against the presence of numerous LIRGs and ULIRGs
in the GRB host population and they also point to lower star formation
rates (see Sect.\,3.2 and Table\,1) which are not consistent with some
conclusions obtained by other groups.  For instance \citet{Berger03}
have recently claimed that 20\% of GRB host galaxies have $SFR \sim
500$\,\Msol\,yr$^{-1}$, which is obviously in disagreement with our
\spi \, results.

Similarly our IRAC data argue very clearly against a population of
sources with massive and evolved stellar populations dominating their
near-IR rest-frame emission.  Assuming a standard conversion between
the mass and the rest-frame near-IR absolute luminosity \citep[i.e.,
  --0.25\,$\ltapp$\,log$_{10}$\,$(\mathcal{M}/L_K)$\,$\ltapp$\,+0.15,
][]{Bell03}, the fluxes or upper limits that we measure at 4.5\mic
translate for most cases into masses
$\mathcal{M}$\,$\ltapp$\,5$\times$10$^{9}$\,\Msol.  The IRAC data are
also consistent with the constraints derived from the optical and the
near-IR published photometry assuming typical SEDs of blue
star-forming galaxies (see Figure\,7), and our results globally agree
with the rather small masses determined by \citet{Chary02} using deep
$K$-band observations of GRB hosts at Keck.  Our interpretation, on
the other hand, strikingly contrasts with the conclusions recently
obtained by \citet{Conselice05} who argue for a trend toward massive
sources at $z$\,$\gtapp$\,1 based on an analysis of the GRB host
morphologies.

Hence the \spi \, view on the GRB-selected galaxies rather suggests
low mass sources characterized by a relatively modest amount of dust
obscuration.  Similar conclusions have already been derived based on
the optical and near-IR properties of these objects
\citep[e.g.,][]{LeFloch03,Courty04,Christensen04}.

\subsection{Implications to GRBs as SFR probes}

Because of their dust-penetrating power and their relation with the
death of young and massive stars it has been argued that the long GRBs
could be used as an unbiased probe of the star formation history of
the Universe. On a pure statistical point of view, their host galaxies
should be therefore representative of the sources producing the bulk
of the stellar mass at high redshift.  Are our results consistent with
this picture\,?

The detection of the cosmic infrared background and the recent surveys
performed at infrared/submillimeter/radio wavelengths have revealed
that a significant fraction of the present-day stellar mass was formed
in the past during short-live and dust-obscured episodes of intense
star formation within infrared-luminous galaxies
\citep[e.g.,][]{Puget96,Blain99,Chary01,Cowie04}.  At $z$\,$\sim$\,1,
these infrared-luminous starbursts (i.e.,
\Lir\,$\geq$\,10$^{11}$\,\Lsol) are responsible for $\sim$70\% of the
star-forming activity in the Universe \citep{LeFloch05} and their
contribution is believed to be even more important at higher redshifts
\citep[e.g.,][]{Blain02,Lagache03}. Using \spi-updated IR galaxy
evolution models \citep[e.g.,][]{Lagache04,Chary04} we estimate that
more than half of the global star-forming activity throughout the
lifetime of the Universe has occured within galaxies characterized by
$f_{24 \mu m}$\,$\gtapp$\,100\muJypa. In this context, the significant
fraction of non-detections in our MIPS data is therefore surprising.

Similarly, the downsizing evolution of the cosmic star formation
history reveals that the bulk of the star-forming activity has been
moving from massive galaxies at high redhifts to low-mass objects in
the present-day Universe \citep[e.g.,][]{Cowie96,Juneau05}.  Should
the long GRBs trace the whole population of distant starbursting
sources, we should therefore detect these bursts preferentially toward
massive systems easily accessible to IRAC.

This apparent discrepancy between the nature of the GRB host galaxies
as a whole and the sources dominating the high-redshift star-forming
activity has already been noted from a comparison between their colors
and luminosities at optical and near-IR wavelengths
\citep[e.g.,][]{LeFloch03} or their star formation rate recovered from
cosmological simulations \citep{Courty04}. For instance GRB hosts
appear blue and optically sub-luminous compared to the massive dusty
starbursts. Very often they also display H$\delta$ in emission
\citep{Djorgovski98,Soderberg04,Prochaska04,Gorosabel05}, which
reveals star formation episodes characterized by younger populations
than those typically observed in the distant luminous-infrared
galaxies \citep{Flores99,Hammer01,Hammer05}. As a result we conclude that the
GRB afterglows, as they are currently selected, can not be considered
as unbiased probes of the integrated activity of star formation in the
high redshift Universe.

\subsection{A bias in the \spi \, sample ?}

As we have described in more detail earlier, our sample consists
mostly of galaxies pinpointed using optical afterglows identified
prior to 2001, when GRB follow-ups were not as prompt and efficient as
they are currently.  This could bias the selection toward the most
luminous transients, and therefore against dust-enshrouded GRBs
occuring within luminous-infrared galaxies. A relevant case
illustrating this hypothesis is given by the dark burst GRB\,970828
and its host galaxy (see Sect.\,\ref{sec:970828}).  The accurate
coordinates of this GRB were determined thanks to the detection of its
radio afterglow, and the absence of optical transient emission was
interpreted as an evidence for dust extinction in the host galaxy
\citep{Djorgovski01a}. The detection of this object at 24\mic clearly
supports this interpretation, and it suggests that a sample of GRB
hosts purely selected with optical afterglows could be biased toward
dust-free sources. In this context, the XRT instrument on-board the
recently-launched {\it Swift\,} satellite is now routinely localizing
X-ray GRB afterglows with an accuracy of $\sim$\,5\arcsec\, on the
sky. In the near-future, the comparison between the properties of GRB
host galaxies selected from optical and X-ray afterglows will provide
better insights into this possible bias affecting our current sample.

On the other hand, we have not detected the two other host galaxies
selected from radio afterglows with no optical transient (i.e., the
hosts of GRB\,981226 and GRB\,990506), which shows that 
not all dark GRBs   originate from  dusty galaxies. A
similar conclusion was reached by \citet{Barnard03} based on the
non-detection of four dark GRB hosts with the SCUBA camera at
850\micpa. In fact, many of these dark bursts could simply be
associated with intrinsically very faint and/or fast-decaying optical
afterglows \citep[e.g.,][]{Fynbo01,Fox03}.

In addition to this observational bias that could affect our
selection, another explanation might be directly related to the
physical properties characterizing the local environments where long
GRBs take place. Based on theoretical simulations and the connection
between GRBs, hypernovae and Type Ic supernovae, it has been argued
that long GRBs should more likely occur within binary systems
\citep{Izzard04,Podsiadlowski04,Mirabel04a,Mirabel04b}, whose
frequency in star-forming regions may vary with redshift.  Furthermore
it has been suggested that long gamma-ray bursts would be more
efficiently produced -- and they would also appear more luminous -- if
they originate from low metallicity progenitors
\citep[e.g.,][]{MacFadyen99,Ramirez_Ruiz02b,Meynet05,Hirschi05}. Massive
stars with metal-poor envelopes keep a high angular momentum in the
latest stage of their evolution, and they are also less subject to
mass loss. After the final collapse, this would favor the formation of
a fast-rotating black hole with accretion of material, that could more
easily lead to a bright GRB event. In this case GRBs would be
preferentially observed in young and chemically-unevolved galaxies,
which would explain this lack of massive and dusty starbursts in our
\spi \, data.

In fact, direct evidence for low metallicity has already been observed
in several GRB host galaxies
\citep{Soderberg04,Prochaska04,Gorosabel05}.  Furthermore, the global
properties of these objects such as their blue colors, their
relatively low luminosities, their Ly$\alpha$ emission and their high
{\it specific\,} star-formation rate
\citep{LeFloch03,Fynbo03,Christensen04} do support a picture where
GRBs occur in low-mass, young and hence metal-poor starbursts.  This
is also corroborated by the cosmological simulations obtained by
\citet{Courty04} who identify the GRB hosts as the most efficient
star-forming objects (i.e., sources with the highest specific
star-formation rate) and not as galaxies with obvious high $SFR$.  In
fact a good illustration of this global property can be given by the
complex environment of the host of GRB\,980613. This burst did not
occur in the region detected by MIPS and harboring therefore the most
intense star-forming activity within the galaxy (see also
\citealt{Hjorth02} for a similar conclusion based on optical HST
data). Note that the same interpretation can be derived from the
characteristics of the GRB\,980425 host galaxy, since the hypernova
SN\,1998bw did not occur in the most active region as revealed by the
very luminous mid-IR point source detected with \spi.

\section{Summary}

We have presented 4.5\micpa/8.0\micpa--IRAC and 24\micpa--MIPS observations
of 16 GRB host galaxies, and our results can be summarized as follows:

\begin{itemize}
\item{} It is now well-established that long GRBs are markers of
  recent bursts of star formation in galaxies.  However most of the
  GRB hosts in our sample were not detected in the rest-frame near-IR
  and mid-IR with \spi, which
argues against a population of sources globally dominated by massive
and IR-luminous starbursts
(\Lir\,$\gtapp$\,5$\times$10$^{11}$\,\Lsol). Current IR galaxy
evolution models indicate that more than half of the integrated
star-forming activity throughout the lifetime of the Universe occured
within $f_{24 \mu m}$\,$\gtapp$\,100\muJy galaxies easily detectable
by MIPS-24\micpa. In this context our results imply that GRBs identified and
localized with the current techniques can not be used as unbiased
probes of the global star formation in the early Universe.
\item{} The detection of the GRB\,970828 host by MIPS at 24\mic
  strongly supports the idea that some of the so-called ``dark
  bursts'' can be explained by the effect of dust extinction within
  their host galaxy. Even though the hosts of the two other dark GRBs
  in our sample (GRB\,981226 and GRB\,990506) were not detected, this
  could indicate that the currently GRB-selected sources are biased
  against dusty starbursts. In the near future the localization of GRB
  hosts using X-ray afterglows detected with the XRT instrument
  on-board {\it Swift\,} will provide new insights on this potential
  bias.

~

\item{} The host of GRB\,010222 that has been claimed to be associated
  with a SCUBA/MAMBO galaxy at $z$\,=\,1.48 is not detected with IRAC
  nor with MIPS, thus bringing strong doubt on the identification of
  this host at long wavelengths.  Similarly our non-detection of the
  host of GRB\,970508 at $z$\,=\,0.84 rules out the proposed
  association of this object with an ultra-luminous infrared galaxy
  seen at 90\mic with ISOPHOT, and the non-detection of the host of
  GRB\,980703 at 24\mic suggests that the radio emission
  previously-detected in this object does not originate from massive
  star formation.

~

\item{} The observations of the host galaxies of GRB\,980425 and
  GRB\,980613 reveal that these bursts did not occur in the regions
  harboring the most active star-forming activity within their
  hosts. This favors a picture in which the production of GRBs does
  not exactly scale with {\it star formation\,} but also depends on
  other parameters that remain to be explored (e.g., age and chemical
  enrichment of the parent progenitor populations, Initial Mass
  Function, fraction of binary systems,...).
\end{itemize}

\acknowledgments {\it Acknowledgments:\,} We thank the funding from
the IRS and the MIPS projects which are supported by NASA through the
Jet Propulsion Laboratory (subcontracts \#1257184 \& \#960785), as
well as the {\it Spitzer\,} Science Center for efficient technical
support.  This work benefited from data obtained with the Multimission
Archive at the Space Telescope Science Institute (MAST) and from
publicly available software with material credited to STScI, which is
operated by the Association of Universities for Research in Astronomy,
Inc., under NASA contract NAS5-26555.  We are also particularly
grateful to the team led by \citet{Holland00a} for publicly providing
reduced HST data from their ``Survey of the Host Galaxies of Gamma-Ray
Bursts''.  We finally thank all our colleagues who shared with us
stimulating discussions related to this work, as well as Ranga-Ram
Chary and our anonymous referee for critical comments on the
manuscript.

\clearpage

\begin{deluxetable}{ccccccccccc}
\tabletypesize{\scriptsize}
\tablenum{1}
\footnotesize
\setlength{\tabcolsep}{0.01in}
\tablecaption{Mid-IR photometry of GRB host galaxies}
\tablewidth{0pc}
\tablehead{
\colhead{Name}  & RA(J2000) & Dec(J2000) & \colhead{f$_{4.5 \mu m}$\,$^{(a)}$} & \colhead{f$_{8 \mu m}$\,$^{(a)}$} & \colhead{f$_{24 \mu m}$\,$^{(a)}$}
 & & \multicolumn{2}{c}{Redshift} &  L$_{\rm IR}$\,$^{(b)}$ & $SFR$\,$_{\rm IR}$\,$^{(c)}$ \\
 & & & \colhead{($\mu$Jy)} & \colhead{($\mu$Jy)} & \colhead{($\mu$Jy)} &  & \multicolumn{2}{c}{--------------------------} & \colhead{(L$_{\sun}$)} & \colhead{(M$_\odot$ yr$^{-1}$)} \\
  &   & &                      &                      &                      &                          & \colhead{z} & \colhead{Reference} &  & }
\startdata
GRB970508 & 06:53:49.45 & +79:16:19.5 &         $<3.0$ & $<16.5$         &       $<83$     & &    0.84 &       1 &   $<1 \times 10^{11}$ & $<17$   \\
GRB970828 & 18:08:31.60 & +59:18:51.5 &    3.7$\pm$0.2 & $<17.8$         &   85$\pm$15     & &    0.96 &       2 &  $1.4^{+2.5}_{-0.8} \times 10^{11}$ &    $24^{+43}_{-14}$   \\
GRB971214 & 11:56:26.40 & +65:12:00.5 &         $<3.6$ & $<17.1$         &       $<92$     & &    3.42 &       3 & $<4.7 \times 10^{13}$ & $<8100$ \\
GRB980326 & 08:36:34.28 & -18:51:23.9 &         $<3.6$ & $<15.1$         &       $<86$     & & \nodata & \nodata &               \nodata & \nodata \\
GRB980329 & 07:02:38.02 & +38:50:44.0 &         $<4.8$ & $<24.4$         &       $<97$     & & \nodata & \nodata &               \nodata & \nodata  \\
GRB980425 & 19:35:03.02 & -52:50:44.8 & 2\,950$\pm$100 & 11\,900$\pm$300 & 27\,300$\pm$200 & &  0.0085 &       4 &     $2 \times 10^{9}$~$^{(d)}$ & 0.4 \\
GRB980519 & 23:22:21.50 & +77:15:43.3 &         $<4.2$ & $<24.4$         &       $<93$     & & \nodata & \nodata &               \nodata & \nodata \\
GRB980613 & 10:17:57.82 & +71:27:25.5 &   36.1$\pm$1.6 & 34.9$\pm$1.7    &  170$\pm$30     & &    1.10 &       5 &    $5^{+9}_{-3} \times 10^{11}$ & $87^{+156}_{-52}$ \\
GRB980703 & 23:59:06.67 & +08:35:07.1 &   11.2$\pm$0.6 & $<23.7$         &       $<83$     & &    0.97 &       6 & $<1.4 \times 10^{11}$ & $<24$ \\
GRB981226 & 23:29:37.21 & -23:55:53.8 &    4.1$\pm$0.2 & $<29.6$         &       $<90$     & & \nodata & \nodata &               \nodata & \nodata \\
GRB990123 & 15:25:30.31 & +44:45:59.2 &         $<3.6$ & $<17.1$         &       $<82$     & &    1.60 &       7 &   $<8 \times 10^{11}$ & $<140$ \\
GRB990308 & 12:23:11.44 & +06:44:05.1 &         $<4.8$ & $<23.7$         &       $<87$     & & \nodata & \nodata &               \nodata & \nodata \\
GRB990506 & 11:54:50,14 & -26:40:35.0 &   1.6~$^{(e)}$ & $<23.7$         &       $<80$     & &    1.31 &       8 & $<4.8 \times 10^{11}$ & $<83$ \\
GRB990510 & 13:38:07.11 & -80:29:48.2 &         $<4.2$ & $<18.4$         &       $<98$     & &    1.62 &       9 & $<9.8 \times 10^{11}$ & $<170$ \\
GRB990705 & 05:09:54.50 & -72:07:53.0 &   17.2$\pm$0.8 & $<17.8$         &  150$\pm$20     & &    0.84 &      10 &  $1.8^{+2.1}_{-0.6} \times 10^{11}$ &     $32^{+37}_{-11}$ \\
GRB010222 & 14:52:12.55 & +43:01:06.3 &         $<3.0$ & $<21.1$         &       $<81$     & &    1.48 &      11 & $<7.4 \times 10^{11}$ & $<130$ 
\enddata
\tablenotetext{(a)}{~All upper limits indicated are 3$\sigma$.} 
\tablenotetext{(b)}{~defined as the energy density integrated between 8\mic and 1000\micpa, and computed assuming  a $\Lambda$CDM cosmology with }
\tablenotetext{\,}{~~H$_0$\,=\,70~km~s$^{-1}$\,Mpc$^{-1}$, $\Omega_m$\,=\,0.3 and $\Omega_{\lambda}\,=\,0.7$.} 
\tablenotetext{(c)}{~assuming $SFR$\,$_{\rm IR}$ (M$_\odot$ yr$^{-1}$) = 1.72\,$\times$\,10$^{-10}$\,L$_{\rm IR}$ (L$_\odot$).} 
\tablenotetext{(d)}{~More than 75\% of the 24\mic monochromatic  luminosity  of this galaxy originates from a single HII region (see Sect.\,4.1). This unusual }
\tablenotetext{\,}{~~property makes the extrapolation to the total infrared highly  uncertain.} 
\tablenotetext{(e)}{~The source is only detected at a $\sim$\,2$\sigma$ level. 
\\
{\bf References:} (1) \citet{Bloom98a} ; (2) \citet{Djorgovski01a} ; (3) \citet{Kulkarni98} ; 
(4) \citet{Tinney98} ; (5) \citet{Djorgovski03} ; (6) \citet{Djorgovski98} ; (7) \citet{Kulkarni99} ; 
(8) \citet{Bloom03} ; (9) \citet{Vreeswijk01} ; (10) \citet{LeFloch02a} ; (11) \citet{Jha01}. 
}
\end{deluxetable}

\clearpage

\begin{figure*}[htpb]
  \epsscale{1.15}
  \plotone{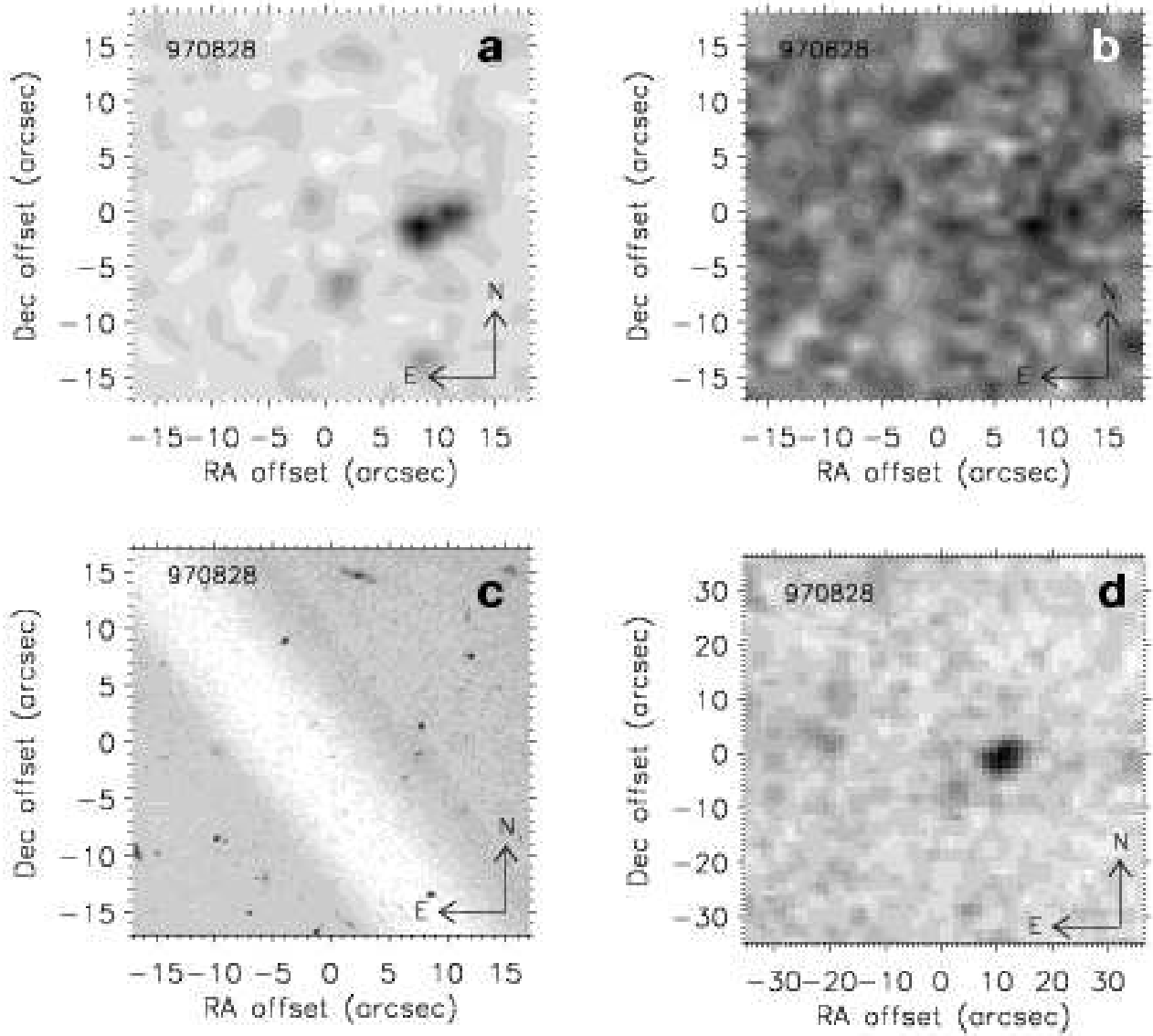}
\caption{Postage stamp mosaic of the area surrounding the GRB\,970828
  host ($z$\,=\,0.96). The location of the GRB is at the center of each image. a)
  Image at 4.5$\mu$m, b) Image at 8.0$\mu$m, c) Optical HST image (see
  text), d) Image at 24$\mu$m. North is to the top and East is to the left.}
\end{figure*}

\clearpage

\begin{figure*}[htpb]
  \epsscale{1.15}
  \plotone{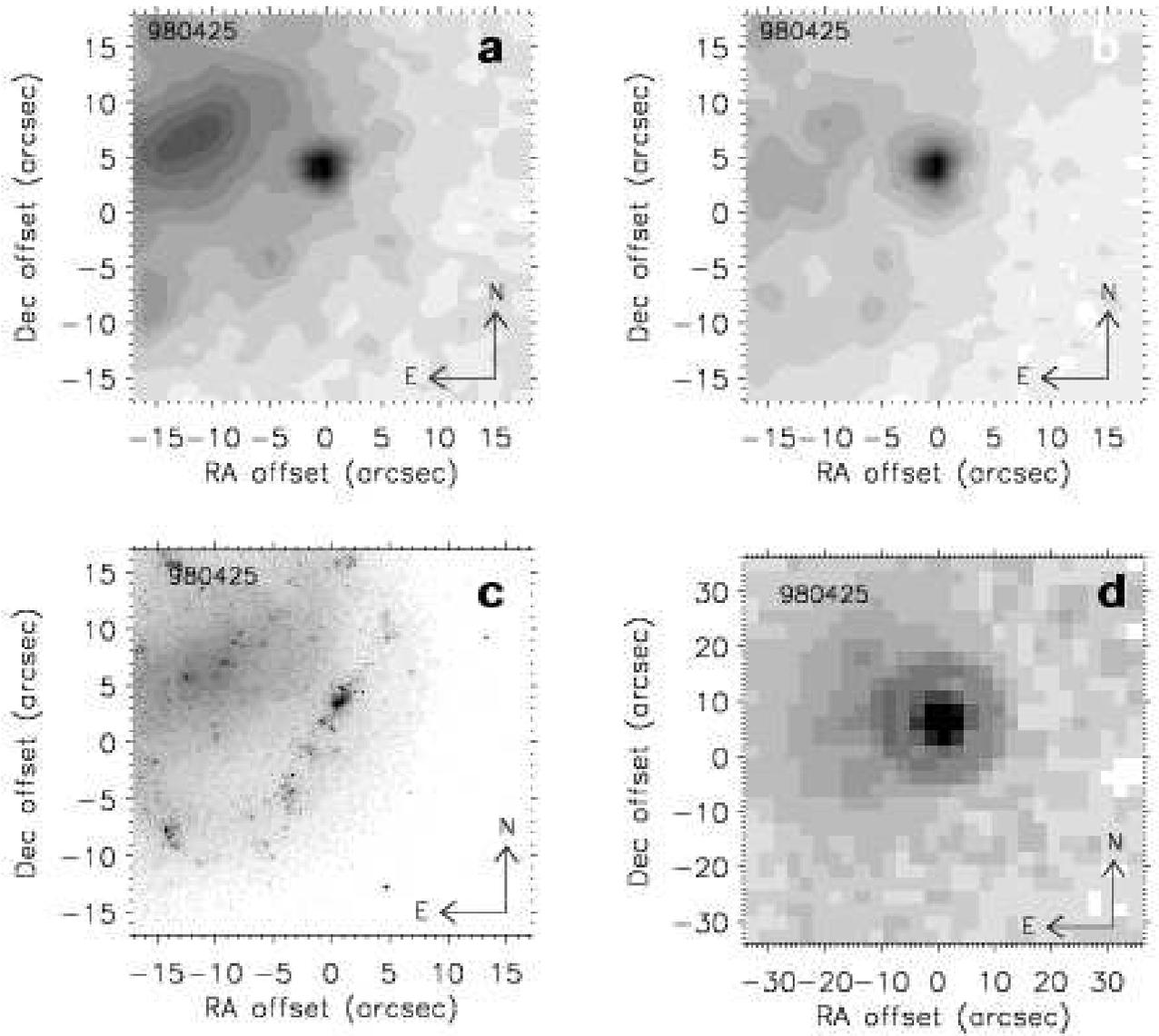}
  \caption{ Postage stamp mosaic of the area surrounding the
    GRB\,980425 host at $z$\,=\,0.0085 (same legend as in Fig.\,1). Note that the bright
point source detected at the \spi \, wavelengths is not exactly  located at the
position of the GRB but rather coincides with the HII region detected 6''
away in the north-west direction.}
\end{figure*}

\clearpage

\begin{figure*}[htpb]
  \epsscale{1.15}
  \plotone{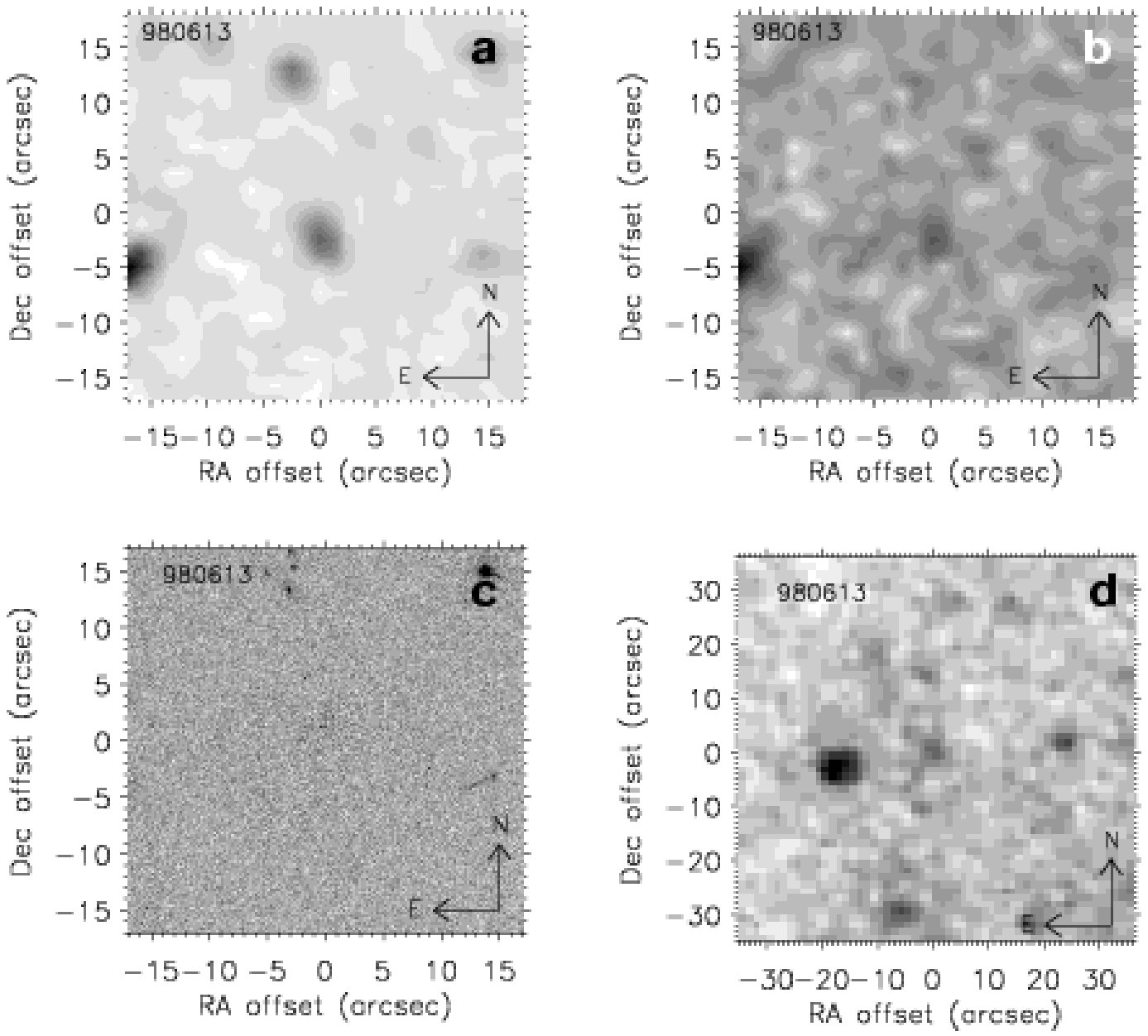}
  \caption{Postage stamp mosaic of the area surrounding the
    GRB\,980613 host at $z$\,=\,1.10 (same legend as in Fig.\,1).} 
\end{figure*}

\clearpage

\begin{figure*}[htpb]
  \epsscale{1.15}
  \plotone{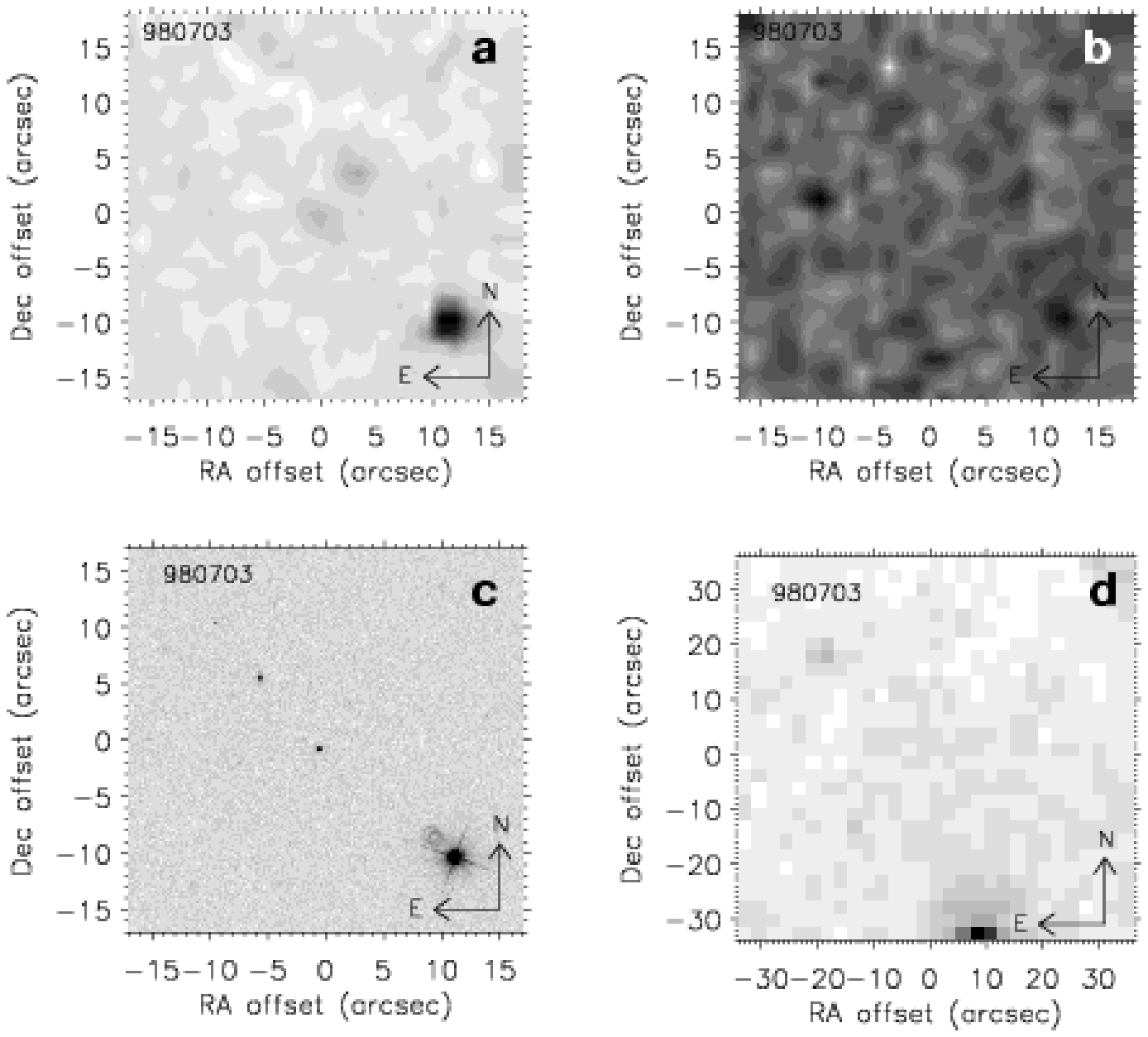}
  \caption{ Postage stamp mosaic of the area surrounding the
    GRB\,980703 host at $z$\,=\,0.97 (same legend as in Fig.\,1). }
\end{figure*}

\clearpage

\begin{figure*}[htpb]
  \epsscale{1.15}
  \plotone{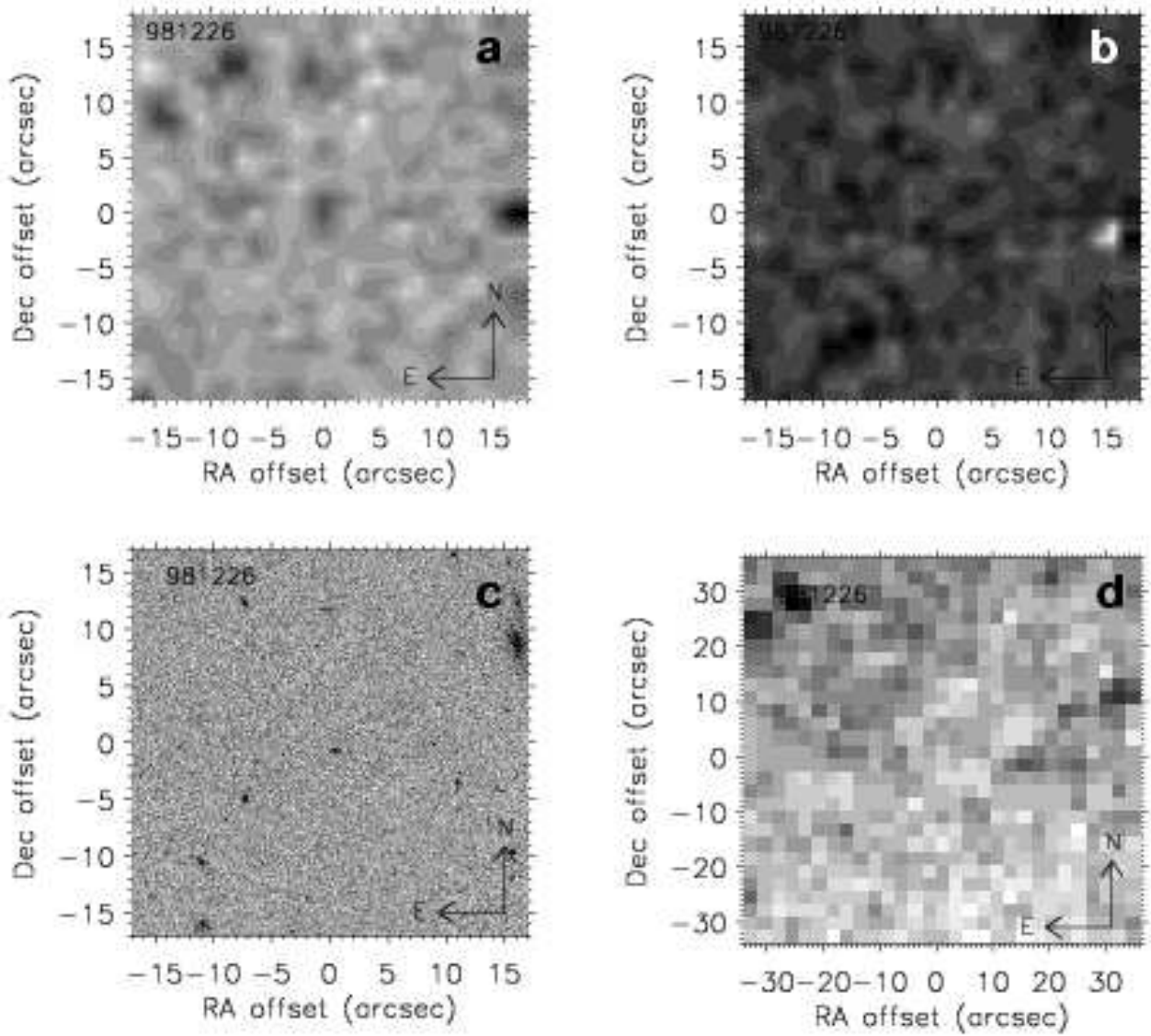}
  \caption{Postage stamp mosaic of the area surrounding the
    GRB\,981226 host (same legend as in Fig.\,1).}
\end{figure*}

\clearpage

\begin{figure*}[htpb]
  \epsscale{1.15}
  \plotone{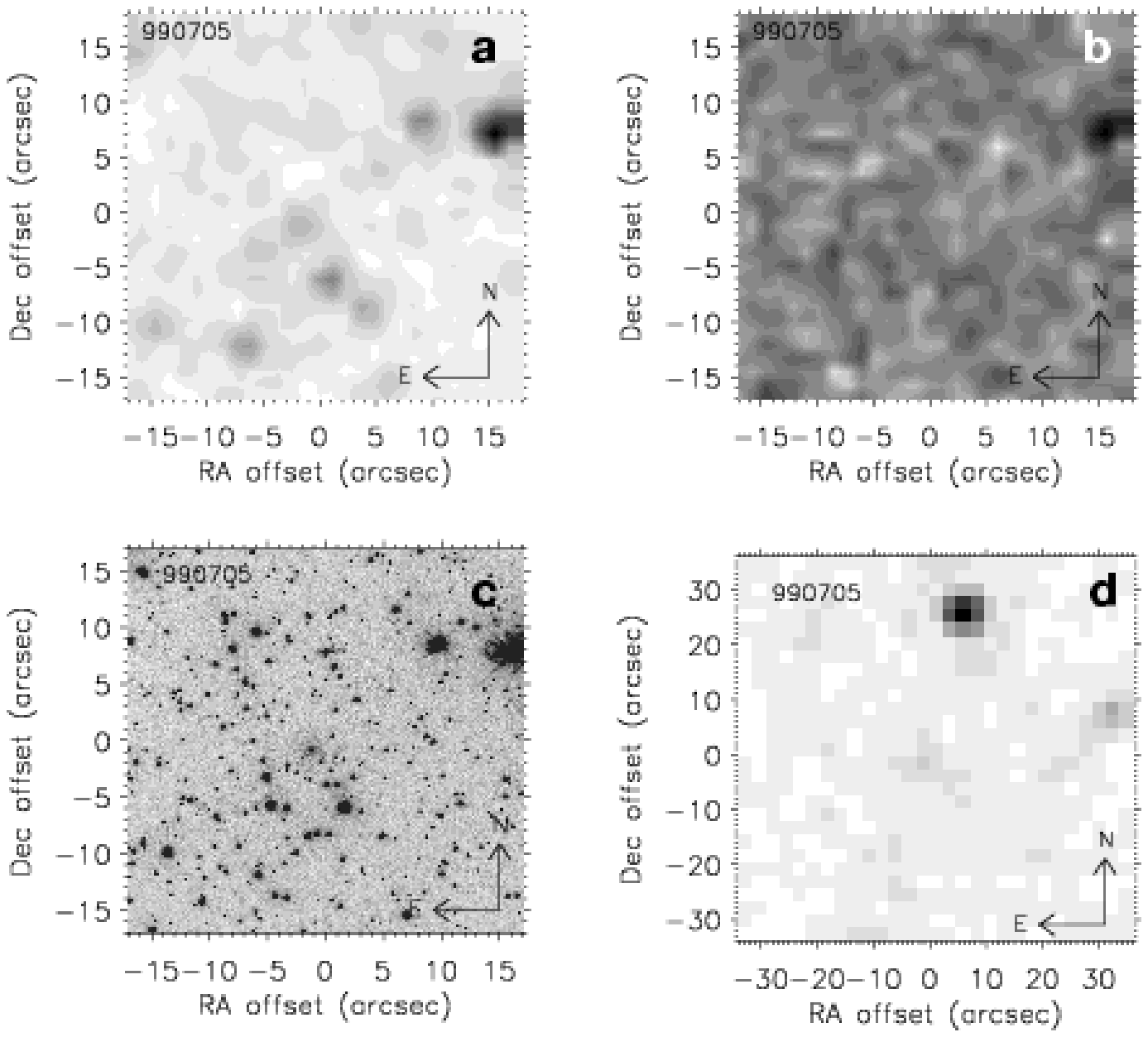}
  \caption{Postage stamp mosaic of the area surrounding the
    GRB\,990705 host at $z$\,=\,0.84 (same legend as in Fig.\,1).}
\end{figure*}

\clearpage
  \epsscale{1.15}
\plotone{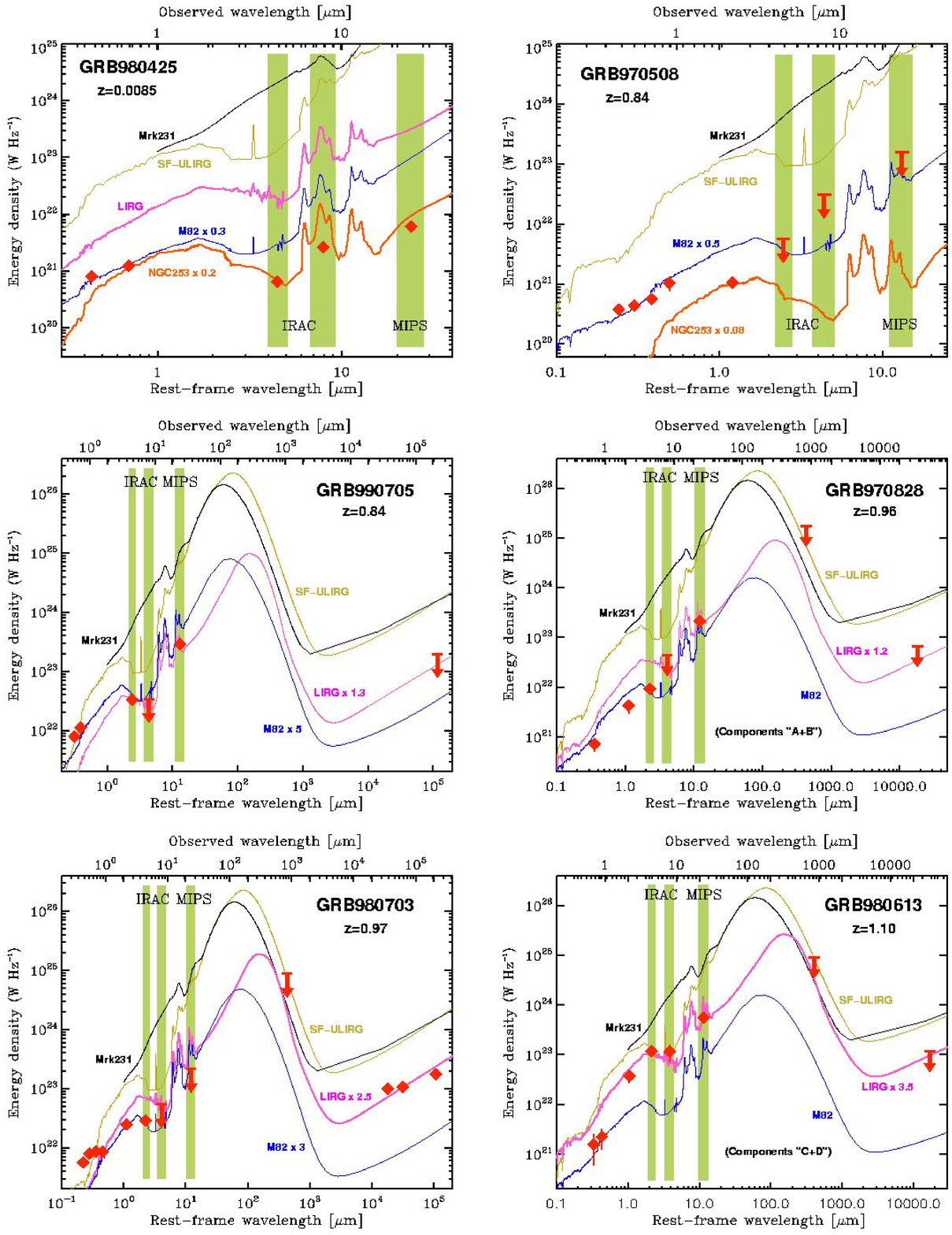}\\
Fig 7a ---

\clearpage

\begin{figure*}
  \epsscale{1.15}
\figurenum{7b}
\plotone{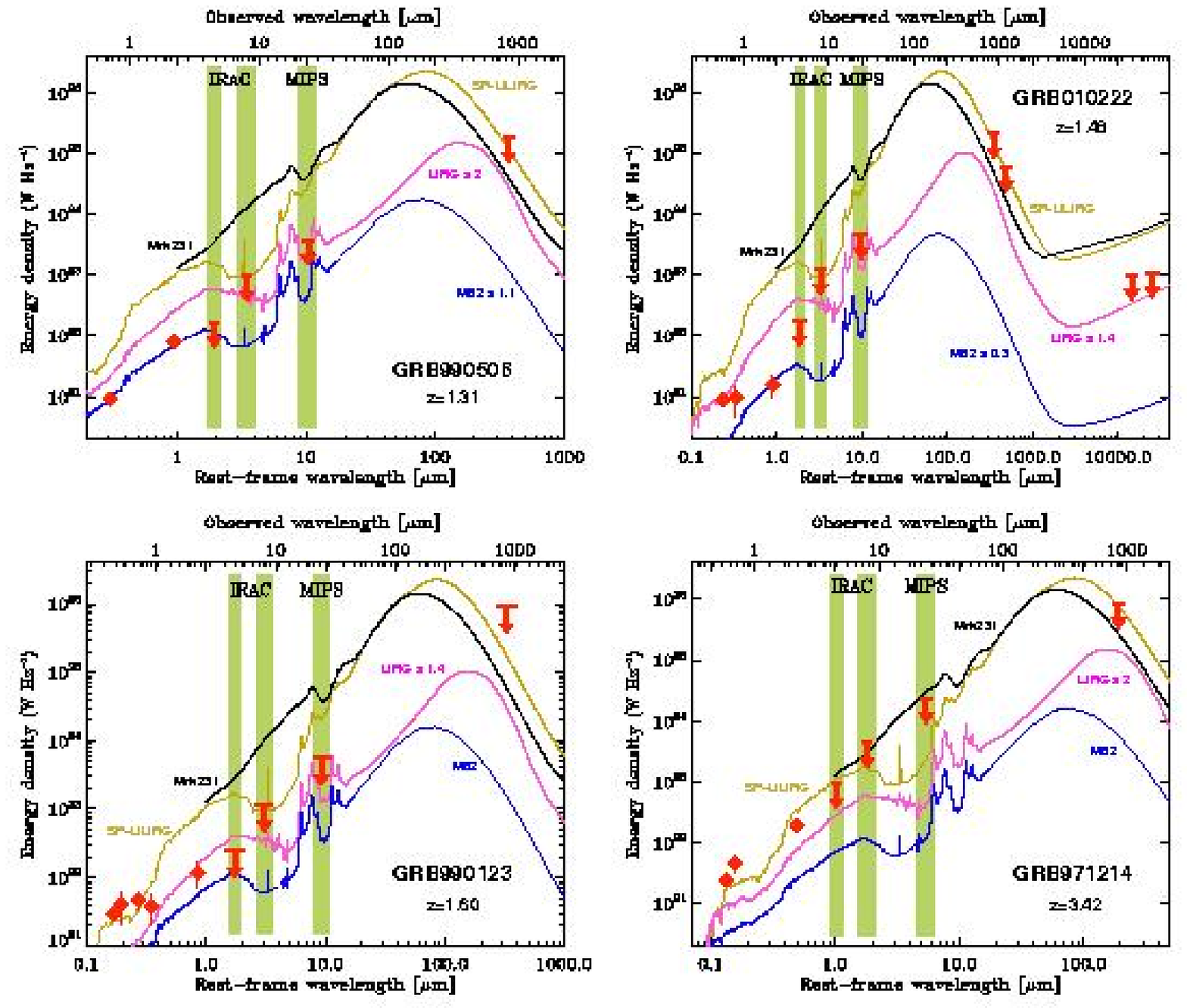}
\caption{Monochromatic luminosities of GRB host galaxies computed from
  the optical to the radio (red filled diamonds, upper limits appear
  as red downward arrows).  Optical, near-IR, submillimeter and radio
  photometry measurements are taken from \citet{Bloom98a},
  \citet{Kulkarni98}, \citet{Bloom99}, \citet{Fruchter99a},
  \citet{Holland99}, \citet{Vreeswijk99}, \citet{Fruchter00b},
  \citet{Fynbo00}, \citet{Berger01b}, \citet{Djorgovski01a},
  \citet{Holland01}, \citet{Sokolov01}, \citet{Chary02},
  \citet{Frail02}, \citet{LeFloch02a}, \citet{Barnard03},
  \citet{Berger03}, \citet{Djorgovski03}, \citet{LeFloch03} and
  \citet{Tanvir04}.  For comparison we also show the 
SEDs of NGC\,253 (orange), M\,82 (blue) and Mrk\,231 (black) as well
as the SED of a starburst-dominated ULIRG with 
L$_{\rm IR}$\,=\,4\,$\times$\,10$^{12}$\,\Lsol \, (light brown, denoted ``SF-ULIRG'')
and the SED of a cold LIRG with L$_{\rm IR}$\,=\,10$^{11}$\,\Lsol \, (purple).
 All these SEDs were corrected for the distance and they are displayed
 in units of luminosity. As indicated in each panel, a scaling factor
 was applied to some of them in order to match the observed GRB host
 photometry at certain wavelengths. The bands where this normalization
 was performed were chosen on a case-by-case basis to better highlight
 the differences in luminosity at the other wavelengths 
 (see text for more detail).  Note that the host of GRB\,980613 is a
 merging system, and the photometry reported in the figure does not
 refer to the region where the GRB occured but only to the component
 detected by \spi \, (see Sect.\,4.6).}  \stepcounter{figure}
\end{figure*}

\clearpage

\begin{figure*}[htpb]
  \epsscale{1.15}
  \plotone{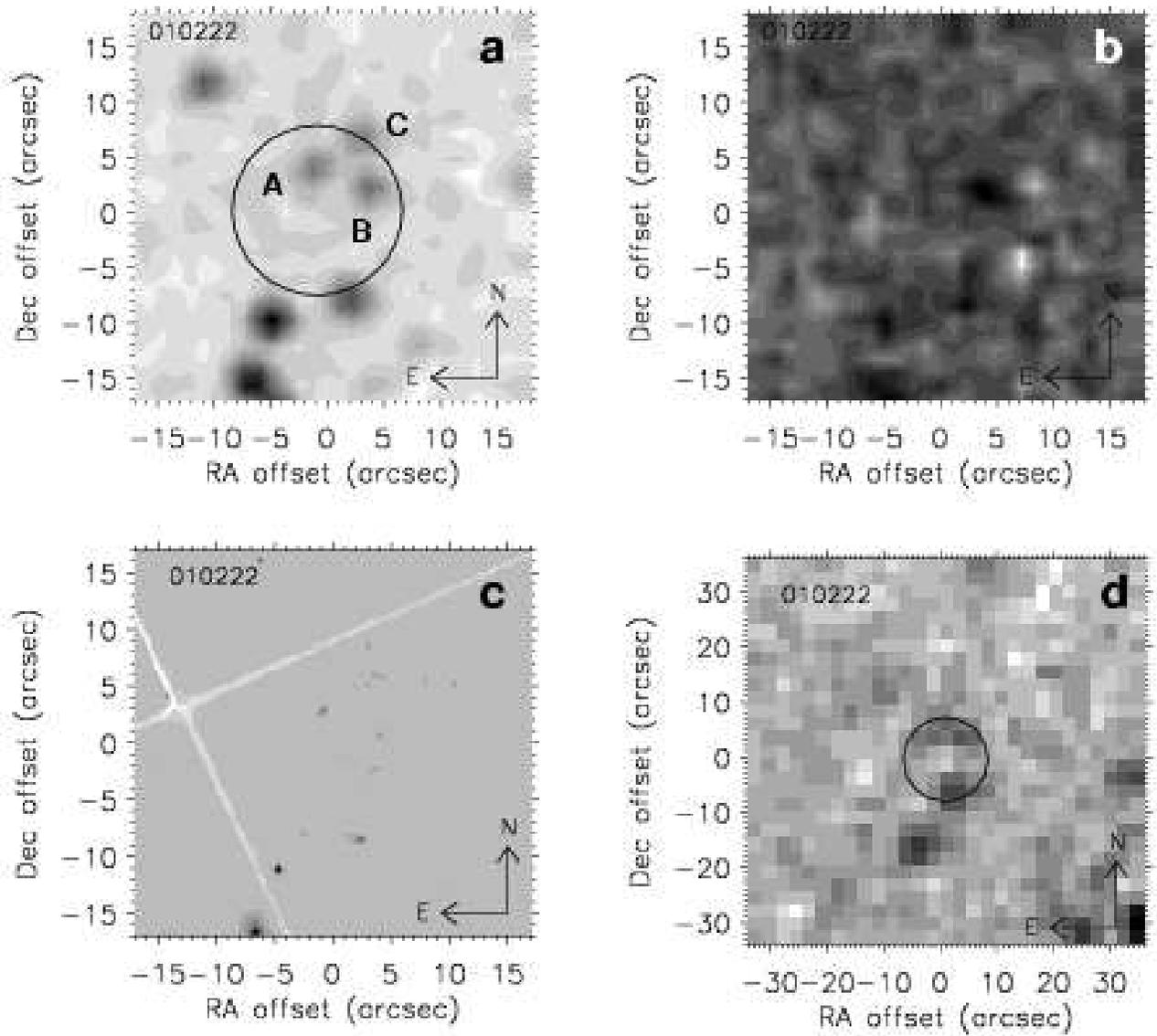}
\caption{Postage stamp mosaic of the area surrounding the GRB\,010222
  host galaxy at $z$\,=\,1.48 (same legend as in Fig.\,1). The 15\arcsec-diameter beam of 
SCUBA centered on the host is displayed in the IRAC 4.5\mic and MIPS 24mic 
images. In panel a), we also indicate the three sources refered to as
galaxies ``A'', ``B'' and ``C'' by \citet{Frail02}.}
\end{figure*}


\begin{thebibliography}{}


\bibitem[{{Appleton} {et~al.}(2004){Appleton}, {Fadda}, {Marleau}, {Frayer},
  {Helou}, {Condon}, {Choi}, {Yan}, {Lacy}, {Wilson}, {Armus}, {Chapman},
  {Fang}, {Heinrichson}, {Im}, {Jannuzi}, {Storrie-Lombardi}, {Shupe},
  {Soifer}, {Squires}, \& {Teplitz}}]{Appleton04}
{Appleton}, P.~N., {Fadda}, D.~T., {Marleau}, F.~R., {et~al.} 2004, \apjs, 154,
  147

\bibitem[{{Armus} {et~al.}(2004){Armus}, {Charmandaris}, {Spoon}, {Houck},
  {Soifer}, {Brandl}, {Appleton}, {Teplitz}, {Higdon}, {Weedman}, {Devost},
  {Morris}, {Uchida}, {van Cleve}, {Barry}, {Sloan}, {Grillmair}, {Burgdorf},
  {Fajardo-Acosta}, {Ingalls}, {Higdon}, {Hao}, {Bernard-Salas}, {Herter},
  {Troeltzsch}, {Unruh}, \& {Winghart}}]{Armus04}
{Armus}, L., {Charmandaris}, V., {Spoon}, H.~W.~W., {et~al.} 2004, \apjs, 154,
  178

\bibitem[{{Barmby} {et~al.}(2004){Barmby}, {Huang}, {Fazio}, {Surace},
  {Arendt}, {Hora}, {Pahre}, {Adelberger}, {Eisenhardt}, {Erb}, {Pettini},
  {Reach}, {Reddy}, {Shapley}, {Steidel}, {Stern}, {Wang}, \&
  {Willner}}]{Barmby04}
{Barmby}, P., {Huang}, J.-S., {Fazio}, G.~G., {et~al.} 2004, \apjs, 154, 97

\bibitem[{{Barnard} {et~al.}(2003){Barnard}, {Blain}, {Tanvir}, {Natarajan},
  {Smith}, {Wijers}, {Kouveliotou}, {Rol}, {Tilanus}, \&
  {Vreeswijk}}]{Barnard03}
{Barnard}, V.~E., {Blain}, A.~W., {Tanvir}, N.~R., {et~al.} 2003, \mnras, 338,
  1

\bibitem[{{Bell} {et~al.}(2003){Bell}, {McIntosh}, {Katz}, \&
  {Weinberg}}]{Bell03}
{Bell}, E.~F., {McIntosh}, D.~H., {Katz}, N., \& {Weinberg}, M.~D. 2003, \apjs,
  149, 289

\bibitem[{{Bell} {et~al.}(2005){Bell}, {Papovich}, {Wolf}, {Le Floc'h},
  {Caldwell}, {Barden}, {Egami}, {McIntosh}, {Meisenheimer}, {P{\'
  e}rez-Gonz{\' a}lez}, {Rieke}, {Rieke}, {Rigby}, \& {Rix}}]{Bell05}
{Bell}, E.~F., {Papovich}, C., {Wolf}, C., {et~al.} 2005, \apj, 625, 23

\bibitem[{{Berger} {et~al.}(2005){Berger}, {Kulkarni}, {Fox}, {Soderberg},
  {Harrison}, {Nakar}, {Kelson}, {Gladders}, {Mulchaey}, {Oemler}, {Dressler},
  {Cenko}, {Price}, {Schmidt}, {Frail}, {Morrell}, {Gonzalez}, {Krzeminski},
  {Sari}, {Gal-Yam}, {Moon}, {Penprase}, {Jayawardhana}, {Scholz}, {Rich},
  {Peterson}, {Anderson}, {McNaught}, {Minezaki}, {Yoshii}, {Cowie}, \&
  {Pimbblet}}]{Berger05}
{Berger}, E., {Kulkarni}, S.~R., {Fox}, D.~B., {et~al.} 2005, \apj, 634, 501

\bibitem[{{Berger} {et~al.}(2003){Berger}, {Cowie}, {Kulkarni}, {Frail},
  {Aussel}, \& {Barger}}]{Berger03}
{Berger}, E., {Cowie}, L.~L., {Kulkarni}, S.~R., {et~al.} 2003, \apj, 588, 99

\bibitem[{{Berger} {et~al.}(2001){Berger}, {Kulkarni}, \& {Frail}}]{Berger01b}
{Berger}, E., {Kulkarni}, S.~R., \& {Frail}, D.~A. 2001, \apj, 560, 652

\bibitem[{{Blain} \& {Natarajan}(2000)}]{Blain00}
{Blain}, A.~W. \& {Natarajan}, P. 2000, \mnras, 312, L35

\bibitem[{{Blain} {et~al.}(1999){Blain}, {Smail}, {Ivison}, \&
  {Kneib}}]{Blain99}
{Blain}, A.~W., {Smail}, I., {Ivison}, R.~J., \& {Kneib}, J.-P. 1999, \mnras,
  302, 632

\bibitem[{{Blain} {et~al.}(2002){Blain}, {Smail}, {Ivison}, {Kneib}, \&
  {Frayer}}]{Blain02}
{Blain}, A.~W., {Smail}, I., {Ivison}, R.~J., {Kneib}, J.-P., \& {Frayer},
  D.~T. 2002, \physrep, 369, 111

\bibitem[{{Bloom} {et~al.}(2003){Bloom}, {Berger}, {Kulkarni}, {Djorgovski}, \&
  {Frail}}]{Bloom03}
{Bloom}, J.~S., {Berger}, E., {Kulkarni}, S.~R., {Djorgovski}, S.~G., \&
  {Frail}, D.~A. 2003, \aj, 125, 999

\bibitem[{{Bloom} {et~al.}(2001{\natexlab{a}}){Bloom}, {Djorgovski}, \&
  {Kulkarni}}]{Bloom01b}
{Bloom}, J.~S., {Djorgovski}, S.~G., \& {Kulkarni}, S.~R. 2001{\natexlab{a}},
  \apj, 554, 678

\bibitem[{{Bloom} {et~al.}(1998){Bloom}, {Djorgovski}, {Kulkarni}, \&
  {Frail}}]{Bloom98a}
{Bloom}, J.~S., {Djorgovski}, S.~G., {Kulkarni}, S.~R., \& {Frail}, D.~A. 1998,
  \apjl, 507, L25

\bibitem[{{Bloom} {et~al.}(2002){Bloom}, {Kulkarni}, \&
  {Djorgovski}}]{Bloom02a}
{Bloom}, J.~S., {Kulkarni}, S.~R., \& {Djorgovski}, S.~G. 2002, \aj, 123, 1111

\bibitem[{{Bloom} {et~al.}(2001){Bloom}, {Kulkarni}, {Galama}, {Frail}, \&
  {Djorgovski}}]{Bloom01c}
{Bloom}, J.~S., {Kulkarni}, S.~R., {Galama}, T.~J., {Frail}, D.~A., \&
  {Djorgovski}, S.~G. 2001, GRB Circular Network, 1134

\bibitem[{{Bloom} {et~al.}(1999){Bloom}, {Odewahn}, {Djorgovski}, {Kulkarni},
  {Harrison}, {Koresko}, {Neugebauer}, {Armus}, {Frail}, {Gal}, {Sari},
  {Squires}, {Illingworth}, {Kelson}, {Chaffee}, {Goodrich}, {Feroci}, {Costa},
  {Piro}, {Frontera}, {Mao}, {Akerlof}, \& {McKay}}]{Bloom99}
{Bloom}, J.~S., {Odewahn}, S.~C., {Djorgovski}, S.~G., {et~al.} 1999, \apjl,
  518, L1

\bibitem[{{Cardelli} {et~al.}(1989){Cardelli}, {Clayton}, \&
  {Mathis}}]{Cardelli89}
{Cardelli}, J.~A., {Clayton}, G.~C., \& {Mathis}, J.~S. 1989, \apj, 345, 245

\bibitem[{{Chapman} {et~al.}(2003){Chapman}, {Blain}, {Ivison}, \&
  {Smail}}]{Chapman03}
{Chapman}, S.~C., {Blain}, A.~W., {Ivison}, R.~J., \& {Smail}, I.~R. 2003,
  \nat, 422, 695

\bibitem[{{Charmandaris} {et~al.}(2004){Charmandaris}, {Uchida}, {Weedman},
  {Herter}, {Houck}, {Teplitz}, {Armus}, {Brandl}, {Higdon}, {Soifer},
  {Appleton}, {van Cleve}, \& {Higdon}}]{Charmandaris04b}
{Charmandaris}, V., {Uchida}, K.~I., {Weedman}, D., {et~al.} 2004, \apjs, 154,
  142

\bibitem[{{Chary} {et~al.}(2002){Chary}, {Becklin}, \& {Armus}}]{Chary02}
{Chary}, R., {Becklin}, E.~E., \& {Armus}, L. 2002, \apj, 566, 229

\bibitem[{{Chary} {et~al.}(2004){Chary}, {Casertano}, {Dickinson}, {Ferguson},
  {Eisenhardt}, {Elbaz}, {Grogin}, {Moustakas}, {Reach}, \& {Yan}}]{Chary04}
{Chary}, R., {Casertano}, S., {Dickinson}, M.~E., {et~al.} 2004, \apjs, 154, 80

\bibitem[{{Chary} \& {Elbaz}(2001)}]{Chary01}
{Chary}, R. \& {Elbaz}, D. 2001, \apj, 556, 562

\bibitem[{{Christensen} {et~al.}(2004){Christensen}, {Hjorth}, \&
  {Gorosabel}}]{Christensen04}
{Christensen}, L., {Hjorth}, J., \& {Gorosabel}, J. 2004, \aap, 425, 913

\bibitem[{{Condon}(1992)}]{Condon92}
{Condon}, J.~J. 1992, \araa, 30, 575

\bibitem[{{Conselice} {et~al.}(2005){Conselice}, {Vreeswijk}, {Fruchter},
  {Levan}, {Kouveliotou}, {Fynbo}, {Gorosabel}, {Tanvir}, \&
  {Thorsett}}]{Conselice05}
{Conselice}, C.~J., {Vreeswijk}, P.~M., {Fruchter}, A.~S., {et~al.} 2005, \apj,
  633, 29

\bibitem[{{Courty} {et~al.}(2004){Courty}, {Bj{\" o}rnsson}, \&
  {Gudmundsson}}]{Courty04}
{Courty}, S., {Bj{\" o}rnsson}, G., \& {Gudmundsson}, E.~H. 2004, \mnras, 354,
  581

\bibitem[{{Cowie} {et~al.}(2004){Cowie}, {Barger}, {Fomalont}, \&
  {Capak}}]{Cowie04}
{Cowie}, L.~L., {Barger}, A.~J., {Fomalont}, E.~B., \& {Capak}, P. 2004, \apjl,
  603, L69

\bibitem[{{Cowie} {et~al.}(1996){Cowie}, {Songaila}, {Hu}, \&
  {Cohen}}]{Cowie96}
{Cowie}, L.~L., {Songaila}, A., {Hu}, E.~M., \& {Cohen}, J.~G. 1996, \aj, 112,
  839

\bibitem[{{Dale} {et~al.}(2005){Dale}, {Bendo}, {Engelbracht}, {Gordon},
  {Regan}, {Armus}, {Cannon}, {Calzetti}, {Draine}, {Helou}, {Joseph},
  {Kennicutt}, {Li}, {Murphy}, {Roussel}, {Walter}, {Hanson}, {Hollenbach},
  {Jarrett}, {Kewley}, {Lamanna}, {Leitherer}, {Meyer}, {Rieke}, {Rieke},
  {Sheth}, {Smith}, \& {Thornley}}]{Dale05}
{Dale}, D.~A., {Bendo}, G.~J., {Engelbracht}, C.~W., {et~al.} 2005, \apj, 633,
  857

\bibitem[{{Dale} {et~al.}(2001){Dale}, {Helou}, {Contursi}, {Silbermann}, \&
  {Kolhatkar}}]{Dale01}
{Dale}, D.~A., {Helou}, G., {Contursi}, A., {Silbermann}, N.~A., \&
  {Kolhatkar}, S. 2001, \apj, 549, 215

\bibitem[{{Dickinson} {et~al.}(2003){Dickinson}, {Papovich}, {Ferguson}, \&
  {Budav{\' a}ri}}]{Dickinson03}
{Dickinson}, M., {Papovich}, C., {Ferguson}, H.~C., \& {Budav{\' a}ri}, T.
  2003, \apj, 587, 25

\bibitem[{{Djorgovski} {et~al.}(2003){Djorgovski}, {Bloom}, \&
  {Kulkarni}}]{Djorgovski03}
{Djorgovski}, S.~G., {Bloom}, J.~S., \& {Kulkarni}, S.~R. 2003, \apjl, 591, L13

\bibitem[{{Djorgovski} {et~al.}(2001){Djorgovski}, {Frail}, {Kulkarni},
  {Bloom}, {Odewahn}, \& {Diercks}}]{Djorgovski01a}
{Djorgovski}, S.~G., {Frail}, D.~A., {Kulkarni}, S.~R., {et~al.} 2001, \apj,
  562, 654

\bibitem[{{Djorgovski} {et~al.}(1998){Djorgovski}, {Kulkarni}, {Bloom},
  {Goodrich}, {Frail}, {Piro}, \& {Palazzi}}]{Djorgovski98}
{Djorgovski}, S.~G., {Kulkarni}, S.~R., {Bloom}, J.~S., {et~al.} 1998, \apjl,
  508, L17

\bibitem[{{Egami} {et~al.}(2004){Egami}, {Dole}, {Huang}, {P{\'
  e}rez-Gonzalez}, {Le Floc'h}, {Papovich}, {Barmby}, {Ivison}, {Serjeant},
  {Mortier}, {Frayer}, {Rigopoulou}, {Lagache}, {Rieke}, {Willner},
  {Alonso-Herrero}, {Bai}, {Engelbracht}, {Fazio}, {Gordon}, {Hines},
  {Misselt}, {Miyazaki}, {Morrison}, {Rieke}, {Rigby}, \& {Wilson}}]{Egami04a}
{Egami}, E., {Dole}, H., {Huang}, J.-S., {et~al.} 2004, \apjs, 154, 130

\bibitem[{{Elbaz} {et~al.}(2002){Elbaz}, {Flores}, {Chanial}, {Mirabel},
  {Sanders}, {Duc}, {Cesarsky}, \& {Aussel}}]{Elbaz02}
{Elbaz}, D., {Flores}, H., {Chanial}, P., {et~al.} 2002, \aap, 381, L1

\bibitem[{{F{\" o}rster Schreiber} {et~al.}(2003){F{\" o}rster Schreiber},
  {Sauvage}, {Charmandaris}, {Laurent}, {Gallais}, {Mirabel}, \&
  {Vigroux}}]{ForsterSchreiber03}
{F{\" o}rster Schreiber}, N.~M., {Sauvage}, M., {Charmandaris}, V., {et~al.}
  2003, \aap, 399, 833

\bibitem[{{Farrah} {et~al.}(2003){Farrah}, {Afonso}, {Efstathiou},
  {Rowan-Robinson}, {Fox}, \& {Clements}}]{Farrah03}
{Farrah}, D., {Afonso}, J., {Efstathiou}, A., {et~al.} 2003, \mnras, 343, 585

\bibitem[{{Fazio} {et~al.}(2004){Fazio}, {Hora}, {Allen}, {Ashby}, {Barmby},
  {Deutsch}, {Huang}, {Kleiner}, {Marengo}, {Megeath}, {Melnick}, {Pahre},
  {Patten}, {Polizotti}, {Smith}, {Taylor}, {Wang}, {Willner}, {Hoffmann},
  {Pipher}, {Forrest}, {McMurty}, {McCreight}, {McKelvey}, {McMurray}, {Koch},
  {Moseley}, {Arendt}, {Mentzell}, {Marx}, {Losch}, {Mayman}, {Eichhorn},
  {Krebs}, {Jhabvala}, {Gezari}, {Fixsen}, {Flores}, {Shakoorzadeh}, {Jungo},
  {Hakun}, {Workman}, {Karpati}, {Kichak}, {Whitley}, {Mann}, {Tollestrup},
  {Eisenhardt}, {Stern}, {Gorjian}, {Bhattacharya}, {Carey}, {Nelson},
  {Glaccum}, {Lacy}, {Lowrance}, {Laine}, {Reach}, {Stauffer}, {Surace},
  {Wilson}, {Wright}, {Hoffman}, {Domingo}, \& {Cohen}}]{Fazio04a}
{Fazio}, G.~G., {Hora}, J.~L., {Allen}, L.~E., {et~al.} 2004, \apjs, 154, 10

\bibitem[{{Flores} {et~al.}(1999){Flores}, {Hammer}, {D{\' e}sert}, {C{\'
  e}sarsky}, {Thuan}, {Crampton}, {Eales}, {Le F{\` e}vre}, {Lilly}, {Omont},
  \& {Elbaz}}]{Flores99}
{Flores}, H., {Hammer}, F., {D{\' e}sert}, F.~X., {et~al.} 1999, \aap, 343, 389

\bibitem[{{Fox} {et~al.}(2003){Fox}, {Price}, {Soderberg}, {Berger},
  {Kulkarni}, {Sari}, {Frail}, {Harrison}, {Yost}, {Matthews}, {Peterson},
  {Tanaka}, {Christiansen}, \& {Moriarty-Schieven}}]{Fox03}
{Fox}, D.~W., {Price}, P.~A., {Soderberg}, A.~M., {et~al.} 2003, \apjl, 586, L5

\bibitem[{{Frail} {et~al.}(2002){Frail}, {Bertoldi}, {Moriarty-Schieven},
  {Berger}, {Price}, {Bloom}, {Sari}, {Kulkarni}, {Gerardy}, {Reichart},
  {Djorgovski}, {Galama}, {Harrison}, {Walter}, {Shepherd}, {Halpern}, {Peck},
  {Menten}, {Yost}, \& {Fox}}]{Frail02}
{Frail}, D.~A., {Bertoldi}, F., {Moriarty-Schieven}, G.~H., {et~al.} 2002,
  \apj, 565, 829

\bibitem[{{Frail} {et~al.}(1999){Frail}, {Kulkarni}, {Bloom}, {Djorgovski},
  {Gorjian}, {Gal}, {Meltzer}, {Sari}, {Chaffee}, {Goodrich}, {Frontera}, \&
  {Costa}}]{Frail99}
{Frail}, D.~A., {Kulkarni}, S.~R., {Bloom}, J.~S., {et~al.} 1999, \apjl, 525,
  L81

\bibitem[{{Franceschini} {et~al.}(2001){Franceschini}, {Aussel}, {Cesarsky},
  {Elbaz}, \& {Fadda}}]{Franceschini01}
{Franceschini}, A., {Aussel}, H., {Cesarsky}, C.~J., {Elbaz}, D., \& {Fadda},
  D. 2001, \aap, 378, 1

\bibitem[{{Franceschini} {et~al.}(2003){Franceschini}, {Berta}, {Rigopoulou},
  {Aussel}, {Cesarsky}, {Elbaz}, {Genzel}, {Moy}, {Oliver}, {Rowan-Robinson},
  \& {Van der Werf}}]{Franceschini03}
{Franceschini}, A., {Berta}, S., {Rigopoulou}, D., {et~al.} 2003, \aap, 403,
  501

\bibitem[{{Frayer} {et~al.}(2004){Frayer}, {Chapman}, {Yan}, {Armus}, {Helou},
  {Fadda}, {Morganti}, {Garrett}, {Appleton}, {Choi}, {Fang}, {Heinrichsen},
  {Im}, {Lacy}, {Marleau}, {Masci}, {Shupe}, {Soifer}, {Squires},
  {Storrie-Lombardi}, {Surace}, {Teplitz}, \& {Wilson}}]{Frayer04}
{Frayer}, D.~T., {Chapman}, S.~C., {Yan}, L., {et~al.} 2004, \apjs, 154, 137

\bibitem[{{Fruchter} {et~al.}(2000{\natexlab{a}}){Fruchter}, {Hook}, \&
  {Pian}}]{Fruchter00c}
{Fruchter}, A., {Hook}, R., \& {Pian}, E. 2000{\natexlab{a}}, GRB Circular
  Network, 757

\bibitem[{{Fruchter} {et~al.}(2001){Fruchter}, {Vreeswijk}, \&
  {Nugent}}]{Fruchter01d}
{Fruchter}, A., {Vreeswijk}, P., \& {Nugent}, P. 2001, GRB Circular Network,
  1029

\bibitem[{{Fruchter}(1999)}]{Fruchter99}
{Fruchter}, A.~S. 1999, \apjl, 512, L1

\bibitem[{{Fruchter} {et~al.}(2000{\natexlab{b}}){Fruchter}, {Pian}, {Gibbons},
  {Thorsett}, {Ferguson}, {Petro}, {Sahu}, {Livio}, {Caraveo}, {Frontera},
  {Kouveliotou}, {Macchetto}, {Palazzi}, {Pedersen}, {Tavani}, \& {van
  Paradijs}}]{Fruchter00b}
{Fruchter}, A.~S., {Pian}, E., {Gibbons}, R., {et~al.} 2000{\natexlab{b}},
  \apj, 545, 664

\bibitem[{{Fruchter} {et~al.}(1999){Fruchter}, {Thorsett}, {Metzger}, {Sahu},
  {Petro}, {Livio}, {Ferguson}, {Pian}, {Hogg}, {Galama}, {Gull},
  {Kouveliotou}, {Macchetto}, {van Paradijs}, {Pedersen}, \&
  {Smette}}]{Fruchter99a}
{Fruchter}, A.~S., {Thorsett}, S.~E., {Metzger}, M.~R., {et~al.} 1999, \apjl,
  519, L13

\bibitem[{{Fynbo} {et~al.}(2003){Fynbo}, {Jakobsson}, {M{\" o}ller}, {Hjorth},
  {Thomsen}, {Andersen}, {Fruchter}, {Gorosabel}, {Holland}, {Ledoux},
  {Pedersen}, {Rhoads}, {Weidinger}, \& {Wijers}}]{Fynbo03}
{Fynbo}, J.~P.~U., {Jakobsson}, P., {M{\" o}ller}, P., {et~al.} 2003, \aap,
  406, L63

\bibitem[{{Fynbo} {et~al.}(2000){Fynbo}, {Holland}, {Andersen}, {Thomsen},
  {Hjorth}, {Bj{\" o}rnsson}, {Jaunsen}, {Natarajan}, \& {Tanvir}}]{Fynbo00}
{Fynbo}, J.~U., {Holland}, S., {Andersen}, M.~I., {et~al.} 2000, \apjl, 542,
  L89

\bibitem[{{Fynbo} {et~al.}(2001){Fynbo}, {Jensen}, {Gorosabel}, {Hjorth},
  {Pedersen}, {M{\o}ller}, {Abbott}, {Castro-Tirado}, {Delgado}, {Greiner},
  {Henden}, {Magazz{\` u}}, {Masetti}, {Merlino}, {Masegosa}, {{\O}stensen},
  {Palazzi}, {Pian}, {Schwarz}, {Cline}, {Guidorzi}, {Goldsten}, {Hurley},
  {Mazets}, {McClanahan}, {Montanari}, {Starr}, \& {Trombka}}]{Fynbo01}
{Fynbo}, J.~U., {Jensen}, B.~L., {Gorosabel}, J., {et~al.} 2001, \aap, 369, 373

\bibitem[{{Galama} {et~al.}(1998){Galama}, {Vreeswijk}, {van Paradijs},
  {Kouveliotou}, {Augusteijn}, {Bohnhardt}, {Brewer}, {Doublier}, {Gonzalez},
  {Leibundgut}, {Lidman}, {Hainaut}, {Patat}, {Heise}, {in 't Zand}, {Hurley},
  {Groot}, {Strom}, {Mazzali}, {Iwamoto}, {Nomoto}, {Umeda}, {Nakamura},
  {Young}, {Suzuki}, {Shigeyama}, {Koshut}, {Kippen}, {Robinson}, {de Wildt},
  {Wijers}, {Tanvir}, {Greiner}, {Pian}, {Palazzi}, {Frontera}, {Masetti},
  {Nicastro}, {Feroci}, {Costa}, {Piro}, {Peterson}, {Tinney}, {Boyle},
  {Cannon}, {Stathakis}, {Sadler}, {Begam}, \& {Ianna}}]{Galama98}
{Galama}, T.~J., {Vreeswijk}, P.~M., {van Paradijs}, J., {et~al.} 1998, \nat,
  395, 67

\bibitem[{{Gorosabel} {et~al.}(2005){Gorosabel}, {P{\'e}rez-Ram{\'{\i}}rez},
  {Sollerman}, {de Ugarte Postigo}, {Fynbo}, {Castro-Tirado}, {Jakobsson},
  {Christensen}, {Hjorth}, {J{\'o}hannesson}, {Guziy}, {Castro Cer{\'o}n},
  {Bj{\"o}rnsson}, {Sokolov}, {Fatkhullin}, \& {Nilsson}}]{Gorosabel05}
{Gorosabel}, J., {P{\'e}rez-Ram{\'{\i}}rez}, D., {Sollerman}, J., {et~al.}
  2005, \aap, 444, 711

\bibitem[{{Gruppioni} {et~al.}(2003){Gruppioni}, {Pozzi}, {Zamorani},
  {Ciliegi}, {Lari}, {Calabrese}, {La Franca}, \& {Matute}}]{Gruppioni03}
{Gruppioni}, C., {Pozzi}, F., {Zamorani}, G., {et~al.} 2003, \mnras, 341, L1

\bibitem[{{Hammer} {et~al.}(2005){Hammer}, {Flores}, {Elbaz}, {Zheng}, {Liang},
  \& {Cesarsky}}]{Hammer05}
{Hammer}, F., {Flores}, H., {Elbaz}, D., {et~al.} 2005, \aap, 430, 115

\bibitem[{{Hammer} {et~al.}(2001){Hammer}, {Gruel}, {Thuan}, {Flores}, \&
  {Infante}}]{Hammer01}
{Hammer}, F., {Gruel}, N., {Thuan}, T.~X., {Flores}, H., \& {Infante}, L. 2001,
  \apj, 550, 570

\bibitem[{{Hanlon} {et~al.}(2000){Hanlon}, {Laureijs}, {Metcalfe}, {McBreen},
  {Altieri}, {Castro-Tirado}, {Claret}, {Costa}, {Delaney}, {Feroci},
  {Frontera}, {Galama}, {Gorosabel}, {Groot}, {Heise}, {Kessler},
  {Kouveliotou}, {Palazzi}, {van Paradijs}, {Piro}, \& {Smith}}]{Hanlon00}
{Hanlon}, L., {Laureijs}, R.~J., {Metcalfe}, L., {et~al.} 2000, \aap, 359, 941

\bibitem[{{Harper} \& {Low}(1971)}]{Harper71}
{Harper}, D.~A. \& {Low}, F.~J. 1971, \apjl, 165, L9+

\bibitem[{{Heger} {et~al.}(2003){Heger}, {Fryer}, {Woosley}, {Langer}, \&
  {Hartmann}}]{Heger03}
{Heger}, A., {Fryer}, C.~L., {Woosley}, S.~E., {Langer}, N., \& {Hartmann},
  D.~H. 2003, \apj, 591, 288

\bibitem[{{Higdon} {et~al.}(2005){Higdon}, {Higdon}, {Weedman}, {Houck}, {Le
  Floc'h}, {Brown}, {Dey}, {Jannuzi}, {Soifer}, \& {Rieke}}]{Higdon05}
{Higdon}, J.~L., {Higdon}, S.~J.~U., {Weedman}, D.~W., {et~al.} 2005, \apj,
  626, 58

\bibitem[{{Hirschi} {et~al.}(2005){Hirschi}, {Meynet}, \& {Maeder}}]{Hirschi05}
{Hirschi}, R., {Meynet}, G., \& {Maeder}, A. 2005, \aap, 443, 581

\bibitem[{{Hjorth} {et~al.}(2003){Hjorth}, {Sollerman}, {M{\o}ller}, {Fynbo},
  {Woosley}, {Kouveliotou}, {Tanvir}, {Greiner}, {Andersen}, {Castro-Tirado},
  {Castro Cer{\' o}n}, {Fruchter}, {Gorosabel}, {Jakobsson}, {Kaper}, {Klose},
  {Masetti}, {Pedersen}, {Pedersen}, {Pian}, {Palazzi}, {Rhoads}, {Rol}, {van
  den Heuvel}, {Vreeswijk}, {Watson}, \& {Wijers}}]{Hjorth03}
{Hjorth}, J., {Sollerman}, J., {M{\o}ller}, P., {et~al.} 2003, \nat, 423, 847

\bibitem[{{Hjorth} {et~al.}(2002){Hjorth}, {Thomsen}, {Nielsen}, {Andersen},
  {Holland}, {Fynbo}, {Pedersen}, {Jaunsen}, {Halpern}, {Fesen}, {Gorosabel},
  {Castro-Tirado}, {McMahon}, {Hoenig}, {Bj{\" o}rnsson}, {Amati}, {Tanvir}, \&
  {Natarajan}}]{Hjorth02}
{Hjorth}, J., {Thomsen}, B., {Nielsen}, S.~R., {et~al.} 2002, \apj, 576, 113

\bibitem[{{Holland} {et~al.}(2000{\natexlab{a}}){Holland}, {Fynbo}, {Thomsen},
  {Andersen}, {Bjornsson}, {Hjorth}, {Jaunsen}, {Natarajan}, \&
  {Tanvir}}]{Holland00a}
{Holland}, S., {Fynbo}, J., {Thomsen}, B., {et~al.} 2000{\natexlab{a}}, GRB
  Circular Network, 698

\bibitem[{{Holland} {et~al.}(2001){Holland}, {Fynbo}, {Hjorth}, {Gorosabel},
  {Pedersen}, {Andersen}, {Dar}, {Thomsen}, {M{\o}ller}, {Bj{\" o}rnsson},
  {Jaunsen}, {Natarajan}, \& {Tanvir}}]{Holland01}
{Holland}, S., {Fynbo}, J.~P.~U., {Hjorth}, J., {et~al.} 2001, \aap, 371, 52

\bibitem[{{Holland} \& {Hjorth}(1999)}]{Holland99}
{Holland}, S. \& {Hjorth}, J. 1999, \aap, 344, L67

\bibitem[{{Holland} {et~al.}(2000{\natexlab{b}}){Holland}, {Thomsen},
  {Andersen}, {Bjornsson}, {Fynbo}, {Hjorth}, {Jaunsen}, {Natarajan}, \&
  {Tanvir}}]{Holland00b}
{Holland}, S., {Thomsen}, B., {Andersen}, M., {et~al.} 2000{\natexlab{b}}, GRB
  Circular Network, 749

\bibitem[{{Holland} {et~al.}(2000{\natexlab{c}}){Holland}, {Thomsen},
  {Andersen}, {Bjornsson}, {Fynbo}, J., {Jaunsen}, {Natarajan}, \&
  {Tanvir}}]{Holland00d}
---. 2000{\natexlab{c}}, GRB Circular Network, 731

\bibitem[{{Houck} {et~al.}(2004){Houck}, {Roellig}, {van Cleve}, {Forrest},
  {Herter}, {Lawrence}, {Matthews}, {Reitsema}, {Soifer}, {Watson}, {Weedman},
  {Huisjen}, {Troeltzsch}, {Barry}, {Bernard-Salas}, {Blacken}, {Brandl},
  {Charmandaris}, {Devost}, {Gull}, {Hall}, {Henderson}, {Higdon}, {Pirger},
  {Schoenwald}, {Sloan}, {Uchida}, {Appleton}, {Armus}, {Burgdorf},
  {Fajardo-Acosta}, {Grillmair}, {Ingalls}, {Morris}, \& {Teplitz}}]{Houck04a}
{Houck}, J.~R., {Roellig}, T.~L., {van Cleve}, J., {et~al.} 2004, \apjs, 154,
  18

\bibitem[{{Huang} {et~al.}(2005){Huang}, {Rigopoulou}, {Willner}, {Papovich},
  {Shu}, {Ashby}, {Barmby}, {Bundy}, {Conselice}, {Egami},
  {P{\'e}rez-Gonz{\'a}lez}, {Rosenberg}, {Smith}, {Wilson}, \&
  {Fazio}}]{Huang05}
{Huang}, J.-S., {Rigopoulou}, D., {Willner}, S.~P., {et~al.} 2005, \apj, 634,
  137

\bibitem[{{Ivison} {et~al.}(2004){Ivison}, {Greve}, {Serjeant}, {Bertoldi},
  {Egami}, {Mortier}, {Alonso-Herrero}, {Barmby}, {Bei}, {Dole}, {Engelbracht},
  {Fazio}, {Frayer}, {Gordon}, {Hines}, {Huang}, {Le Floc'h}, {Misselt},
  {Miyazaki}, {Morrison}, {Papovich}, {P{\' e}rez-Gonz{\' a}lez}, {Rieke},
  {Rieke}, {Rigby}, {Rigopoulou}, {Smail}, {Wilson}, \& {Willner}}]{Ivison04}
{Ivison}, R.~J., {Greve}, T.~R., {Serjeant}, S., {et~al.} 2004, \apjs, 154, 124

\bibitem[{{Izzard} {et~al.}(2004){Izzard}, {Ramirez-Ruiz}, \&
  {Tout}}]{Izzard04}
{Izzard}, R.~G., {Ramirez-Ruiz}, E., \& {Tout}, C.~A. 2004, \mnras, 348, 1215

\bibitem[{{Jaunsen} {et~al.}(2003){Jaunsen}, {Andersen}, {Hjorth}, {Fynbo},
  {Holland}, {Thomsen}, {Gorosabel}, {Schaefer}, {Bj{\" o}rnsson}, {Natarajan},
  \& {Tanvir}}]{Jaunsen03}
{Jaunsen}, A.~O., {Andersen}, M.~I., {Hjorth}, J., {et~al.} 2003, \aap, 402,
  125

\bibitem[{{Jha} {et~al.}(2001){Jha}, {Pahre}, {Garnavich}, {Calkins},
  {Kilgard}, {Matheson}, {McDowell}, {Roll}, \& {Stanek}}]{Jha01}
{Jha}, S., {Pahre}, M.~A., {Garnavich}, P.~M., {et~al.} 2001, \apjl, 554, L155

\bibitem[{{Juneau} {et~al.}(2005){Juneau}, {Glazebrook}, {Crampton},
  {McCarthy}, {Savaglio}, {Abraham}, {Carlberg}, {Chen}, {Le Borgne}, {Marzke},
  {Roth}, {J{\o}rgensen}, {Hook}, \& {Murowinski}}]{Juneau05}
{Juneau}, S., {Glazebrook}, K., {Crampton}, D., {et~al.} 2005, \apjl, 619, L135

\bibitem[{{Kasliwal} {et~al.}(2005){Kasliwal}, {Charmandaris}, {Weedman},
  {Houck}, {Le Floc'h}, {Higdon}, {Armus}, \& {Teplitz}}]{Kasliwal05}
{Kasliwal}, M.~M., {Charmandaris}, V., {Weedman}, D., {et~al.} 2005, \apjl,
  634, L1

\bibitem[{{Kawai} {et~al.}(2005){Kawai}, {Yamada}, {Kosugi}, {Hattori}, \&
  {Aoki}}]{Kawai05}
{Kawai}, N., {Yamada}, T., {Kosugi}, G., {Hattori}, T., \& {Aoki}, K. 2005, GRB
  Circular Network, 3937, 1

\bibitem[{{Kennicutt}(1998)}]{Kennicutt98}
{Kennicutt}, R.~C. 1998, \araa, 36, 189

\bibitem[{{Kouveliotou} {et~al.}(1993){Kouveliotou}, {Meegan}, {Fishman},
  {Bhat}, {Briggs}, {Koshut}, {Paciesas}, \& {Pendleton}}]{Kouveliotou93}
{Kouveliotou}, C., {Meegan}, C.~A., {Fishman}, G.~J., {et~al.} 1993, \apjl,
  413, L101

\bibitem[{{Kulkarni} {et~al.}(1998){Kulkarni}, {Djorgoski}, {Ramaprakash},
  {Goodrich}, {Bloom}, {Adelberger}, {Kundic}, {Lubin}, {Frail}, {Frontera},
  {Feroci}, {Nicastro}, {Barth}, {Davis}, {Filippenko}, \&
  {Newman}}]{Kulkarni98}
{Kulkarni}, S.~R., {Djorgoski}, S.~G., {Ramaprakash}, A.~N., {et~al.} 1998,
  \nat, 393, 35

\bibitem[{{Kulkarni} {et~al.}(1999){Kulkarni}, {Djorgovski}, {Odewahn},
  {Bloom}, {Gal}, {Koresko}, {Harrison}, {Lubin}, {Armus}, {Sari},
  {Illingworth}, {Kelson}, {Magee}, {van Dokkum}, {Frail}, {Mulchaey},
  {Malkan}, {McClean}, {Teplitz}, {Koerner}, {Kirkpatrick}, {Kobayashi},
  {Yadigaroglu}, {Halpern}, {Piran}, {Goodrich}, {Chaffee}, {Feroci}, \&
  {Costa}}]{Kulkarni99}
{Kulkarni}, S.~R., {Djorgovski}, S.~G., {Odewahn}, S.~C., {et~al.} 1999, \nat,
  398, 389

\bibitem[{{Lagache} {et~al.}(2003){Lagache}, {Dole}, \& {Puget}}]{Lagache03}
{Lagache}, G., {Dole}, H., \& {Puget}, J.-L. 2003, \mnras, 338, 555

\bibitem[{{Lagache} {et~al.}(2004){Lagache}, {Dole}, {Puget}, {P{\'
  e}rez-Gonz{\' a}lez}, {Le Floc'h}, {Rieke}, {Papovich}, {Egami},
  {Alonso-Herrero}, {Engelbracht}, {Gordon}, {Misselt}, \&
  {Morrison}}]{Lagache04}
{Lagache}, G., {Dole}, H., {Puget}, J.-L., {et~al.} 2004, \apjs, 154, 112

\bibitem[{{Lamb} \& {Reichart}(2000)}]{Lamb00}
{Lamb}, D.~Q. \& {Reichart}, D.~E. 2000, \apj, 536, 1

\bibitem[{{Le~Floc'h} {et~al.}(2003){Le~Floc'h}, {Duc}, {Mirabel}, {Sanders},
  {Bosch}, {Diaz}, {Donzelli}, {Rodrigues}, {Courvoisier}, {Greiner},
  {Mereghetti}, {Melnick}, {Maza}, \& {Minniti}}]{LeFloch03}
{Le~Floc'h}, E., {Duc}, P.-A., {Mirabel}, I.~F., {et~al.} 2003, \aap, 400, 499

\bibitem[{{Le~Floc'h} {et~al.}(2002){Le~Floc'h}, {Duc}, {Mirabel}, {Sanders},
  {Bosch}, {Rodrigues}, {Courvoisier}, {Mereghetti}, \& {Melnick}}]{LeFloch02a}
---. 2002, \apjl, 581, L81

\bibitem[{{Le Floc'h} {et~al.}(2005){Le Floc'h}, {Papovich}, {Dole}, {Bell},
  {Lagache}, {Rieke}, {Egami}, {P{\'e}rez-Gonz{\'a}lez}, {Alonso-Herrero},
  {Rieke}, {Blaylock}, {Engelbracht}, {Gordon}, {Hines}, {Misselt}, {Morrison},
  \& {Mould}}]{LeFloch05}
{Le Floc'h}, E., {Papovich}, C., {Dole}, H., {et~al.} 2005, \apj, 632, 169

\bibitem[{{MacFadyen} \& {Woosley}(1999)}]{MacFadyen99}
{MacFadyen}, A.~I. \& {Woosley}, S.~E. 1999, \apj, 524, 262

\bibitem[{{Malesani} {et~al.}(2004){Malesani}, {Tagliaferri}, {Chincarini},
  {Covino}, {Della Valle}, {Fugazza}, {Mazzali}, {Zerbi}, {D'Avanzo},
  {Kalogerakos}, {Simoncelli}, {Antonelli}, {Burderi}, {Campana}, {Cucchiara},
  {Fiore}, {Ghirlanda}, {Goldoni}, {G{\" o}tz}, {Mereghetti}, {Mirabel},
  {Romano}, {Stella}, {Minezaki}, {Yoshii}, \& {Nomoto}}]{Malesani04}
{Malesani}, D., {Tagliaferri}, G., {Chincarini}, G., {et~al.} 2004, \apjl, 609,
  L5

\bibitem[{{Melbourne} {et~al.}(2005){Melbourne}, {Koo}, \& {Le
  Floc'h}}]{Melbourne05}
{Melbourne}, J., {Koo}, D.~C., \& {Le Floc'h}, E. 2005, \apjl, 632, L65

\bibitem[{{Meynet} \& {Maeder}(2005)}]{Meynet05}
{Meynet}, G. \& {Maeder}, A. 2005, \aap, 429, 581

\bibitem[{{Mirabel}(2004{\natexlab{a}})}]{Mirabel04a}
{Mirabel}, I.~F. 2004{\natexlab{a}}, in ESA SP-552: 5th INTEGRAL Workshop on
  the INTEGRAL Universe, 175 (astro--ph/0405433)

\bibitem[{{Mirabel}(2004{\natexlab{b}})}]{Mirabel04b}
{Mirabel}, I.~F. 2004{\natexlab{b}}, in Revista Mexicana de Astronomia y
  Astrofisica Conference Series, 14 (astro--ph/0405256)

\bibitem[{{Mirabel} {et~al.}(2000){Mirabel}, {Sanders}, \& {Le
  Floc'h}}]{Mirabel00}
{Mirabel}, I.~F., {Sanders}, D.~B., \& {Le Floc'h}, E. 2000, in ``Cosmic
  Evolution and Galaxy Formation''. ASP Conf. Series., Vol. 215, Eds. Franco
  J., Terlevich E., Lopez-Cruz O., Aretxaga I. (astro-ph/0004022)

\bibitem[{{Odewahn} {et~al.}(1998){Odewahn}, {Djorgovski}, {Kulkarni},
  {Dickinson}, {Frail}, {Ramaprakash}, {Bloom}, {Adelberger}, {Halpern},
  {Helfand}, {Bahcall}, {Goodrich}, {Frontera}, {Feroci}, {Piro}, \&
  {Costa}}]{Odewahn98}
{Odewahn}, S.~C., {Djorgovski}, S.~G., {Kulkarni}, S.~R., {et~al.} 1998, \apjl,
  509, L5

\bibitem[{{Piro} {et~al.}(2000){Piro}, {Garmire}, {Garcia}, {Stratta}, {Costa},
  {Feroci}, {M{\' e}sz{\' a}ros}, {Vietri}, {Bradt}, {Frail}, {Frontera},
  {Halpern}, {Heise}, {Hurley}, {Kawai}, {Kippen}, {Marshall}, {Murakami},
  {Sokolov}, {Takeshima}, \& {Yoshida}}]{Piro00}
{Piro}, L., {Garmire}, G., {Garcia}, M., {et~al.} 2000, Science, 290, 955

\bibitem[{{Podsiadlowski} {et~al.}(2004){Podsiadlowski}, {Mazzali}, {Nomoto},
  {Lazzati}, \& {Cappellaro}}]{Podsiadlowski04}
{Podsiadlowski}, P., {Mazzali}, P.~A., {Nomoto}, K., {Lazzati}, D., \&
  {Cappellaro}, E. 2004, \apjl, 607, L17

\bibitem[{{Prochaska} {et~al.}(2004){Prochaska}, {Bloom}, {Chen}, {Hurley},
  {Melbourne}, {Dressler}, {Graham}, {Osip}, \& {Vacca}}]{Prochaska04}
{Prochaska}, J.~X., {Bloom}, J.~S., {Chen}, H.-W., {et~al.} 2004, \apj, 611,
  200

\bibitem[{{Puget} {et~al.}(1996){Puget}, {Abergel}, {Bernard}, {Boulanger},
  {Burton}, {Desert}, \& {Hartmann}}]{Puget96}
{Puget}, J.-L., {Abergel}, A., {Bernard}, J.-P., {et~al.} 1996, \aap, 308, L5

\bibitem[{{Ramirez-Ruiz} {et~al.}(2002){Ramirez-Ruiz}, {Lazzati}, \&
  {Blain}}]{Ramirez_Ruiz02b}
{Ramirez-Ruiz}, E., {Lazzati}, D., \& {Blain}, A.~W. 2002, \apjl, 565, L9

\bibitem[{{Reeves} {et~al.}(2002){Reeves}, {Watson}, {Osborne}, {Pounds},
  {O'Brien}, {Short}, {Turner}, {Watson}, {Mason}, {Ehle}, \&
  {Schartel}}]{Reeves02}
{Reeves}, J.~N., {Watson}, D., {Osborne}, J.~P., {et~al.} 2002, \nat, 416, 512

\bibitem[{{Richards}(2000)}]{Richards00}
{Richards}, E.~A. 2000, \apj, 533, 611

\bibitem[{{Rieke} {et~al.}(2004){Rieke}, {Young}, {Engelbracht}, {Kelly},
  {Low}, {Haller}, {Beeman}, {Gordon}, {Stansberry}, {Misselt}, {Cadien},
  {Morrison}, {Rivlis}, {Latter}, {Noriega-Crespo}, {Padgett}, {Stapelfeldt},
  {Hines}, {Egami}, {Muzerolle}, {Alonso-Herrero}, {Blaylock}, {Dole}, {Hinz},
  {Le Floc'h}, {Papovich}, {P{\' e}rez-Gonz{\' a}lez}, {Smith}, {Su},
  {Bennett}, {Frayer}, {Henderson}, {Lu}, {Masci}, {Pesenson}, {Rebull}, {Rho},
  {Keene}, {Stolovy}, {Wachter}, {Wheaton}, {Werner}, \& {Richards}}]{Rieke04}
{Rieke}, G.~H., {Young}, E.~T., {Engelbracht}, C.~W., {et~al.} 2004, \apjs,
  154, 25

\bibitem[{{Roussel} {et~al.}(2001){Roussel}, {Sauvage}, {Vigroux}, \&
  {Bosma}}]{Roussel01}
{Roussel}, H., {Sauvage}, M., {Vigroux}, L., \& {Bosma}, A. 2001, \aap, 372,
  427

\bibitem[{{Sanders} \& {Mirabel}(1996)}]{Sanders96}
{Sanders}, D.~B. \& {Mirabel}, I.~F. 1996, \araa, 34, 749

\bibitem[{{Saracco} {et~al.}(2001){Saracco}, {Chincarini}, {Covino},
  {Ghisellini}, {Longhetti}, {Zerbi}, {Lazzati}, \& {Severgnini}}]{Saracco01a}
{Saracco}, P., {Chincarini}, G., {Covino}, S., {et~al.} 2001, GRB Circular
  Network, 1032

\bibitem[{{Schlegel} {et~al.}(1998){Schlegel}, {Finkbeiner}, \&
  {Davis}}]{Schlegel98}
{Schlegel}, D.~J., {Finkbeiner}, D.~P., \& {Davis}, M. 1998, \apj, 500, 525

\bibitem[{{Silva} {et~al.}(1998){Silva}, {Granato}, {Bressan}, \&
  {Danese}}]{Silva98}
{Silva}, L., {Granato}, G.~L., {Bressan}, A., \& {Danese}, L. 1998, \apj, 509,
  103

\bibitem[{{Smail} {et~al.}(2004){Smail}, {Chapman}, {Blain}, \&
  {Ivison}}]{Smail04}
{Smail}, I., {Chapman}, S.~C., {Blain}, A.~W., \& {Ivison}, R.~J. 2004, \apj,
  616, 71

\bibitem[{{Soderberg} {et~al.}(2004){Soderberg}, {Kulkarni}, {Berger}, {Fox},
  {Price}, {Yost}, {Hunt}, {Frail}, {Walker}, {Hamuy}, {Shectman}, {Halpern},
  \& {Mirabal}}]{Soderberg04}
{Soderberg}, A.~M., {Kulkarni}, S.~R., {Berger}, E., {et~al.} 2004, \apj, 606,
  994

\bibitem[{{Sokolov} {et~al.}(2001){Sokolov}, {Fatkhullin}, {Castro-Tirado},
  {Fruchter}, {Komarova}, {Kasimova}, {Dodonov}, {Afanasiev}, \&
  {Moiseev}}]{Sokolov01}
{Sokolov}, V.~V., {Fatkhullin}, T.~A., {Castro-Tirado}, A.~J., {et~al.} 2001,
  \aap, 372, 438

\bibitem[{{Spergel} {et~al.}(2003){Spergel}, {Verde}, {Peiris}, {Komatsu},
  {Nolta}, {Bennett}, {Halpern}, {Hinshaw}, {Jarosik}, {Kogut}, {Limon},
  {Meyer}, {Page}, {Tucker}, {Weiland}, {Wollack}, \& {Wright}}]{Spergel03}
{Spergel}, D.~N., {Verde}, L., {Peiris}, H.~V., {et~al.} 2003, \apjs, 148, 175

\bibitem[{{Spoon} {et~al.}(2004){Spoon}, {Armus}, {Cami}, {Tielens}, {Chiar},
  {Peeters}, {Keane}, {Charmandaris}, {Appleton}, {Teplitz}, \&
  {Burgdorf}}]{Spoon04}
{Spoon}, H.~W.~W., {Armus}, L., {Cami}, J., {et~al.} 2004, \apjs, 154, 184

\bibitem[{{Stanek} {et~al.}(2003){Stanek}, {Matheson}, {Garnavich}, {Martini},
  {Berlind}, {Caldwell}, {Challis}, {Brown}, {Schild}, {Krisciunas}, {Calkins},
  {Lee}, {Hathi}, {Jansen}, {Windhorst}, {Echevarria}, {Eisenstein}, {Pindor},
  {Olszewski}, {Harding}, {Holland}, \& {Bersier}}]{Stanek03}
{Stanek}, K.~Z., {Matheson}, T., {Garnavich}, P.~M., {et~al.} 2003, \apjl, 591,
  L17

\bibitem[{{Stetson}(1987)}]{Stetson87}
{Stetson}, P.~B. 1987, \pasp, 99, 191

\bibitem[{{Tanvir} {et~al.}(2004){Tanvir}, {Barnard}, {Blain}, {Fruchter},
  {Kouveliotou}, {Natarajan}, {Ramirez-Ruiz}, {Rol}, {Smith}, {Tilanus}, \&
  {Wijers}}]{Tanvir04}
{Tanvir}, N.~R., {Barnard}, V.~E., {Blain}, A.~W., {et~al.} 2004, \mnras, 352,
  1073

\bibitem[{{Taylor} {et~al.}(2000){Taylor}, {Bloom}, {Frail}, {Kulkarni},
  {Djorgovski}, \& {Jacoby}}]{Taylor00}
{Taylor}, G.~B., {Bloom}, J.~S., {Frail}, D.~A., {et~al.} 2000, \apjl, 537, L17

\bibitem[{{Tinney} {et~al.}(1998){Tinney}, {Stathakis}, {Cannon}, {Galama},
  {Wieringa}, {Frail}, {Kulkarni}, {Higdon}, {Wark}, \& {Bloom}}]{Tinney98}
{Tinney}, C., {Stathakis}, R., {Cannon}, R., {et~al.} 1998, \iaucirc, 6896, 1

\bibitem[{{Venemans} \& {Blain}(2001)}]{Venemans01}
{Venemans}, B.~P. \& {Blain}, A.~W. 2001, \mnras, 325, 1477

\bibitem[{{Vreeswijk} {et~al.}(2001){Vreeswijk}, {Fruchter}, {Kaper}, {Rol},
  {Galama}, {van Paradijs}, {Kouveliotou}, {Wijers}, {Pian}, {Palazzi},
  {Masetti}, {Frontera}, {Savaglio}, {Reinsch}, {Hessman}, {Beuermann},
  {Nicklas}, \& {van den Heuvel}}]{Vreeswijk01}
{Vreeswijk}, P.~M., {Fruchter}, A., {Kaper}, L., {et~al.} 2001, \apj, 546, 672

\bibitem[{{Vreeswijk} {et~al.}(1999){Vreeswijk}, {Galama}, {Owens},
  {Oosterbroek}, {Geballe}, {van Paradijs}, {Groot}, {Kouveliotou}, {Koshut},
  {Tanvir}, {Wijers}, {Pian}, {Palazzi}, {Frontera}, {Masetti}, {Robinson},
  {Briggs}, {in 't Zand}, {Heise}, {Piro}, {Costa}, {Feroci}, {Antonelli},
  {Hurley}, {Greiner}, {Smith}, {Levine}, {Lipkin}, {Leibowitz}, {Lidman},
  {Pizzella}, {B{\" o}hnhardt}, {Doublier}, {Chaty}, {Smail}, {Blain}, {Hough},
  {Young}, \& {Suntzeff}}]{Vreeswijk99}
{Vreeswijk}, P.~M., {Galama}, T.~J., {Owens}, A., {et~al.} 1999, \apj, 523, 171

\bibitem[{{Weedman} {et~al.}(2005){Weedman}, {Hao}, {Higdon}, {Devost}, {Wu},
  {Charmandaris}, {Brandl}, {Bass}, \& {Houck}}]{Weedman05}
{Weedman}, D.~W., {Hao}, L., {Higdon}, S.~J.~U., {et~al.} 2005, \apj, 633, 706

\bibitem[{{Werner} {et~al.}(2004){Werner}, {Roellig}, {Low}, {Rieke}, {Rieke},
  {Hoffmann}, {Young}, {Houck}, {Brandl}, {Fazio}, {Hora}, {Gehrz}, {Helou},
  {Soifer}, {Stauffer}, {Keene}, {Eisenhardt}, {Gallagher}, {Gautier}, {Irace},
  {Lawrence}, {Simmons}, {Van Cleve}, {Jura}, {Wright}, \&
  {Cruikshank}}]{Werner04}
{Werner}, M.~W., {Roellig}, T.~L., {Low}, F.~J., {et~al.} 2004, \apjs, 154, 1

\bibitem[{{Wijers} {et~al.}(1998){Wijers}, {Bloom}, {Bagla}, \&
  {Natarajan}}]{Wijers98}
{Wijers}, R.~A.~M.~J., {Bloom}, J.~S., {Bagla}, J.~S., \& {Natarajan}, P. 1998,
  \mnras, 294, L13

\bibitem[{{Woosley}(1993)}]{Woosley93}
{Woosley}, S.~E. 1993, \apj, 405, 273

\bibitem[{{Zheng} {et~al.}(2004){Zheng}, {Hammer}, {Flores}, {Ass{\' e}mat}, \&
  {Pelat}}]{Zheng04}
{Zheng}, X.~Z., {Hammer}, F., {Flores}, H., {Ass{\' e}mat}, F., \& {Pelat}, D.
  2004, \aap, 421, 847

\end{thebibliography}
\end{document}